\documentclass[12pt]{article}
\usepackage{graphicx}

\begin{document}

\title{Violations of the Leggett-Garg inequality for coherent and cat states}

\author{Hiroo Azuma${}^{1,}$\thanks{Email: hiroo.azuma@m3.dion.ne.jp}
\ \ 
and
\ \ 
Masashi Ban${}^{2,}$\thanks{Email: m.ban@phys.ocha.ac.jp}
\\
\\
{\small ${}^{1}$Nisshin-scientia Co., Ltd.,}\\
{\small 8F Omori Belport B, 6-26-2 MinamiOhi, Shinagawa-ku, Tokyo 140-0013, Japan}\\
{\small ${}^{2}$Graduate School of Humanities and Sciences, Ochanomizu University,}\\
{\small 2-1-1 Ohtsuka, Bunkyo-ku, Tokyo 112-8610, Japan}
}

\date{\today}

\maketitle

\begin{abstract}
We show that in some cases the coherent state can have a larger violation of the Leggett-Garg inequality (LGI) than the cat state
by numerical calculations.
To achieve this result,
we consider the LGI of the cavity mode weakly coupled to a zero-temperature environment as a practical instance of the physical system.
We assume that the bosonic mode undergoes dissipation because of an interaction with the environment but is not affected by dephasing.
Solving the master equation exactly,
we derive an explicit form of the violation of the inequality for both systems prepared initially in the coherent state $|\alpha\rangle$
and the cat state $(|\alpha\rangle+|-\alpha\rangle)$.
For the evaluation of the inequality,
we choose the displaced parity operators characterized by a complex number $\beta$.
We look for the optimum parameter $\beta$ that lets the upper bound of the inequality be maximum numerically.
Contrary to our expectations,
the coherent state occasionally exhibits quantum quality more strongly than the cat state for the upper bound of the violation of the LGI
in a specific range of three equally spaced measurement times (spacing $\tau$).
Moreover, as we let $\tau$ approach zero,
the optimized parameter $\beta$ diverges and the LGI reveals intense singularity.
\end{abstract}

\section{\label{section-introduction}Introduction}
A violation of Bell's inequality illustrates the fact that quantum mechanics is essentially incompatible with classical mechanics \cite{Bell1964}.
To compute Bell's inequality,
first we prepare states of two qubits spatially separated,
second observe them independently,
and third evaluate expectation values by taking averages of the set of the measurements.
The violation of Bell's inequality tells us that quantum mechanics is able to exhibit strong correlations which cannot be explained by the hidden variable theory.

The Leggett-Garg inequality (LGI) was proposed by Leggett and Garg in 1985 for testing
whether or not
macroscopic coherence could be realized in the laboratory \cite{Leggett1985,Brukner2004,Emary2014}.
When we consider Bell's inequality,
we pay attention to the correlation between the two spatially separated qubits.
By contrast, although the LGI is an analogue of Bell's one,
it examines a correlation of measurements at two different times for a single particle,
and thus we can regard it as temporal Bell's inequality.

To derive this inequality,
Leggett and Garg introduced the following two assumptions.
The first one is ``macroscopic realism''.
This assumption requires that results of observation are determined by hidden variables that are attributes of the observed system.
Moreover,
we postulate that the hidden variables are independent of the observation itself.
The second one is ``non-invasive measurability''.
This premise implies that a result of measurement at time $t_{2}$ is independent of a result of previous measurement at time $t_{1}(<t_{2})$.
Thus, we can specify the state of the system uniquely without disturbing the system.

In general,
classical mechanics satisfies these two assumptions,
so that dynamics of all systems that evolves according to the classical theory does not violate the LGI.
However,
the two assumptions do not always hold in the quantum mechanics.
Thus,
it is possible that the dynamics of a quantum system violates the LGI.
In fact,
concrete examples of quantum systems which violate the inequality have been already found theoretically
\cite{Huelga1995,Chevalier2009,Chen2014,Lobejko2015,Friedenberger2017,Azuma2018,Thenabadu2019}.
Recently,
experimental demonstrations of the violation of the LGI have been reported \cite{Palacios-Laloy2010,Goggin2011,Knee2012}.

In the current paper,
we investigate the LGI for a cavity mode weakly coupled to a zero-temperature environment.
We make the boson system undergo dissipation due to the interaction with the environment
but we assume that it does not receive dephasing.
Because there is no thermal fluctuation of the external reservoir,
we can solve the master equation of the boson system exactly in a closed-form expression \cite{Walls1994,Barnett1997}.

In the present paper,
we focus on cases where initial states of the system are given by the coherent state $|\alpha\rangle$
and the cat state $(|\alpha\rangle+|-\alpha\rangle)$
and compare their violations of the LGI.
The cat state is also called the even coherent state
and we can consider it to be a special one of the Yurke-Stoler coherent states \cite{Yurke1986,Schleich1991,Buzek1992,Kim1992}.

In order to evaluate the LGI,
selection of operators for dichotomic observation is important.
In the derivation of the inequality,
we performs measurements of the dichotomic variable at three equally spaced times,
$t_{1}=0$, $t_{2}=\tau$, and $t_{3}=2\tau$, where $\tau>0$.
In the present paper,
we choose the displaced parity operators to measure the state of the boson system \cite{Bishop1994,Banaszek1998,Banaszek1999,Kim2000,Jeong2003}.
These operators are characterized by an arbitrary complex number $\beta=re^{i\theta}$.
Thus,
adjusting the two real parameters, $r$ and $\theta$,
we can vary the violation of the LGI.

In the current paper,
we concentrate on a maximization problem,
where we look for the optimum values of the parameters, $\theta$ and $r$,
for maximizing the violation.
Because the LGI reflects the macroscopic coherence of the system of interest,
we can regard its violation as a quantity of the quantum feature of the system.
One of the motivations of the present paper is to clarify which state exhibits a characteristic of quantum nature more strongly,
the coherent state or the cat state.

We obtain the following result:
the violation of the inequality for the coherent state is larger than that for the cat state
if we let the time difference $\tau$ of the LGI be a value belonging to a specific range.
This result is unexpected because the cat state is a superposition of two different coherent states, $|\alpha\rangle$ and $|-\alpha\rangle$,
and we can consider intuitively that the cat state $(|\alpha\rangle+|-\alpha\rangle)$ exhibits the characteristic of quantum nature more intensely
than the ordinary coherent state $|\alpha\rangle$.
This counter intuitive fact is novel and interesting.

Moreover,
if we let the time difference $\tau$ approach zero,
the parameter $\beta$ diverges to infinity as $|\beta|=r\to\infty$ for the optimized displaced parity operators
and the maximized violation of the inequality converges to $3/2$ but not unity,
so that the LGI shows strong singularity.
Because such behaviour of the violation does not occur in the inequality for a two-level system
(i.e. using projection operators as observables,
the upper bound of the inequality for the two-level system gets closer to unity as the time difference $\tau$ approaches zero),
we can guess that the origin of the singularity is the infiniteness of the dimension of the Hilbert space for the boson system.

Here,
we mention previous works.
In the current paper,
we study the LGI for a cavity mode interacting with a zero-temperature environment.
In Ref.~\cite{Chen2014},
Chen {\it et al}. investigated the LGI of a qubit coupled to a
zero-temperature environment.
In Refs.~\cite{Lobejko2015,Friedenberger2017}, the LGI of a qubit interacting with a thermal environment was studied.
In Ref.~\cite{Thenabadu2019}, Thenabadu and Reid discussed the violation of the LGI of the cat state of the boson system
but they did not consider the interaction with an environment.

In Ref.~\cite{Lambert2010a},
two inequalities inspired by the LGI were derived to distinguish quantum from classical transport through nanostructures.
In Refs.~\cite{Lambert2010b,Chen2012},
the extended LGI proposed in Ref.~\cite{Lambert2010a} was utilized
on a double quantum dot and a multiple-quantum-well structure.
In
Ref.~\cite{Li2012},
new practical quantum witnesses to verify quantum coherence for complex systems were proposed and they were regarded as refined ones compared with the LGI.
In Ref.~\cite{Emary2012},
the violation of the LGI in electronic Mach-Zehnder interferometers was investigated.

The current paper is organized as follows.
In Sect.~\ref{section-reviews-master-equation-Leggett-Garg-inequality},
we review the master equation of the boson system,
the LGI, and the displaced parity operators.
In Sect.~\ref{section-LG-equation-for-coherent-state},
we derive the explicit form of the LGI for the system prepared initially in the coherent state.
In Sect.~\ref{section-LG-equation-for-CAT-state},
we derive the rigorous form of the inequality with assuming that the system is initially in the cat state.
In Sect.~\ref{section-initial-coherent-state-numerical-analyses},
we estimate the violation of the inequality numerically in the case where the system is initialized in the coherent state $|\alpha\rangle$.
In Sect.~\ref{section-initial-CAT-state-numerical-analyses},
we evaluate the violation numerically in the case where the system is initially put in the cat state $(|\alpha\rangle+|-\alpha\rangle)$.
In Sects.~\ref{section-initial-coherent-state-numerical-analyses} and \ref{section-initial-CAT-state-numerical-analyses},
we perform optimization of the displaced parity operators.
In Sect.~\ref{section-coherent-CAT-states-numerical-analyses},
we compare the maximized violations of the inequality for the two initial states, the coherent and cat states.
In Sect.~\ref{section-discussion}, we give brief discussion.
In Appendices~\ref{section-appendix-A} and \ref{section-appendix-B},
we give explicit mathematical forms that appear in Sects.~\ref{section-LG-equation-for-coherent-state} and \ref{section-LG-equation-for-CAT-state},
respectively.

\section{\label{section-reviews-master-equation-Leggett-Garg-inequality}
Reviews of the master equation for the boson system, the LGI, and the displaced parity operators}
In the current section, first of all,
we review the master equation for the boson system and its exact solution \cite{Walls1994,Barnett1997}.
Next, we introduce the LGI \cite{Leggett1985,Brukner2004,Emary2014}
and the displaced parity operators \cite{Bishop1994,Banaszek1998,Banaszek1999,Kim2000,Jeong2003}.

The master equation of the boson system weakly coupled to a zero-temperature environment is given by
\begin{equation}
\dot{\rho}(t)=-i[H,\rho(t)]+\Gamma[2a\rho(t)a^{\dagger}-a^{\dagger}a\rho(t)-\rho(t)a^{\dagger}a],
\label{master-equation-0}
\end{equation}
\begin{equation}
H=\omega a^{\dagger}a,
\end{equation}
where we put $\hbar=1$, and $\Gamma$ and $\omega$ represent the spontaneous emission rate and the angular velocity, respectively.
We assume that $\Gamma$ and $\omega$ are real.
Here, we consider how to solve Eq.~(\ref{master-equation-0}) rigorously.

Taking the interaction picture,
we simplify the master equation~(\ref{master-equation-0}) as
\begin{equation}
\dot{\rho}(t)=\Gamma[2a\rho(t)a^{\dagger}-a^{\dagger}a\rho(t)-\rho(t)a^{\dagger}a].
\label{master-equation-1}
\end{equation}
We introduce superoperators $\hat{J}$ and $\hat{L}$,
where we put hats on the symbols to emphasize that they are superoperators,
as follows:
\begin{equation}
\hat{J}\rho(t)=a\rho(t)a^{\dagger},
\end{equation}
\begin{equation}
\hat{L}\rho(t)=-(1/2)[a^{\dagger}a\rho(t)+\rho(t)a^{\dagger}a].
\end{equation}
Then,
the master equation~(\ref{master-equation-1}) can be rewritten in the form,
\begin{equation}
\dot{\rho}(t)=2\Gamma(\hat{J}+\hat{L})\rho(t).
\label{master-equation-operator-form}
\end{equation}

Using a commutation relation
$[\hat{J},\hat{L}]=-\hat{J}$,
we obtain a formal solution of Eq.~(\ref{master-equation-operator-form}),
\begin{eqnarray}
\rho(t)
&=&
\exp[2\Gamma t(\hat{J}+\hat{L})]\rho(0) \nonumber \\
&=&
\exp(2\Gamma t\hat{L})\exp\{[1-\exp(-2\Gamma t)]\hat{J}\}\rho(0).
\end{eqnarray}
Now, we substitute an operator $v(0)=|\alpha\rangle\langle \beta|$ for $\rho(0)$,
where $|\alpha\rangle$ and $|\beta\rangle$ are coherent states and $\alpha$ and $\beta$ are arbitrary complex numbers.
Then, $v(t)$ is given by
\begin{eqnarray}
v(t)
&=&
\exp\{-(1/2)(|\alpha|^{2}+|\beta|^{2}-2\alpha\beta^{*})[1-\exp(-2\Gamma t)]\} \nonumber \\
&&
\times
|\alpha\exp(-\Gamma t)\rangle\langle\beta\exp(-\Gamma t)|.
\label{v-time-evolution}
\end{eqnarray}
Here, we change the interaction picture of Eq.~(\ref{v-time-evolution}) into the Schr{\"{o}}dinger picture.
Finally, we attain
\begin{eqnarray}
v(t)
&=&
\exp\{-(1/2)(|\alpha|^{2}+|\beta|^{2}-2\alpha\beta^{*})[1-\exp(-2\Gamma t)]\} \nonumber \\
&&
\times
|\alpha\exp(-i\Omega t)\rangle\langle\beta\exp(-i\Omega t)|,
\label{time-evolution-alpha-beta-0}
\end{eqnarray}
where
$\Omega=\omega-i\Gamma$.

We pay attention to the following facts.
If we put $v(0)=|\alpha\rangle\langle\beta|$ and $\tilde{v}(0)=|\beta\rangle\langle\alpha|$,
$\tilde{v}(t)^{\dagger}=v(t)$ holds.
Moreover,
if we set $\rho(0)=|\alpha\rangle\langle\alpha|$,
we obtain
\\
$\rho(t)=|\alpha\exp(-i\Omega t)\rangle\langle\alpha\exp(-i\Omega t)|$.

The LGI is defined as follows.
For the sake of simplicity,
we assume that the dichotomic observables are given by projection operators.
We consider an operator $\hat{O}$ whose eigenvalues are equal to $\pm 1$.
To emphasize that it is an operator,
we put a hat on the symbol.
Next, we introduce equally spaced three times,
$t_{1}=0$, $t_{2}=\tau$, and $t_{3}=2\tau$, where $\tau>0$.
We describe an observed value of the measurement with $\hat{O}$ at time $t_{1}$ as $O_{1}$.
Obviously, $O_{1}=\pm 1$ holds.
We write a probability that observed values at times $t_{1}$ and $t_{2}$ are equal to $O_{1}$ and $O_{2}$ respectively as $P_{21}(O_{1},O_{2})$.
We define the correlation function $C_{21}$ as
\begin{equation}
C_{21}
=
\sum_{O_{1},O_{2}\in\{-1,+1\}}O_{2}O_{1}P_{21}(O_{1},O_{2}).
\end{equation}
Then, the LGI is given by
\begin{equation}
K_{3}=C_{21}+C_{32}-C_{31},
\label{definition-Leggett-Garg-inequality-0}
\end{equation}
\begin{equation}
-3\leq K_{3}\leq 1.
\end{equation}

We define projection operators which are called the displaced parity operators as
\begin{eqnarray}
\Pi^{(+)}(\beta)
&=
&
D(\beta)\sum_{n=0}^{\infty}|2n\rangle\langle 2n|D^{\dagger}(\beta), \nonumber \\
\Pi^{(-)}(\beta)
&=
&
D(\beta)\sum_{n=0}^{\infty}|2n+1\rangle\langle 2n+1|D^{\dagger}(\beta),
\label{displaced-parity-operator-definition-0}
\end{eqnarray}
where
\begin{equation}
D(\beta)=\exp(\beta a^{\dagger}-\beta^{*}a).
\end{equation}
In the present paper,
hereafter,
we choose the following operator for the orthogonal measurement of the LGI,
\begin{equation}
\hat{O}=\Pi^{(+)}(\beta)-\Pi^{(-)}(\beta).
\end{equation}

\section{\label{section-LG-equation-for-coherent-state}
The explicit form of the LGI for the system initially prepared in the coherent state}
In the present section,
putting a coherent state $|\alpha\rangle$ as the initial state,
we derive a closed-form expression of the LGI.

The probability that we obtain $O_{1}=1$ for the measurement of the initial state $|\alpha\rangle$ at time $t_{1}=0$ is given by
\begin{eqnarray}
P_{1+}
&=&
\langle\alpha|\Pi^{(+)}(\beta)|\alpha\rangle \nonumber \\
&=&
\exp(-|\alpha-\beta|^{2})
\cosh|\alpha-\beta|^{2}.
\end{eqnarray}
Then, the state collapses from $|\alpha\rangle$ to $w_{1+}(0)/P_{1+}$,
where $w_{1+}(0)$ is given by
\begin{equation}
w_{1+}(0)
=
\Pi^{(+)}(\beta)|\alpha\rangle\langle\alpha|\Pi^{(+)}(\beta),
\label{w1plus0-0}
\end{equation}
\begin{equation}
\Pi^{(+)}(\beta)|\alpha\rangle
=
(1/2)[
|\alpha\rangle
+
\exp(\beta^{*}\alpha-\beta\alpha^{*})|2\beta-\alpha\rangle
].
\label{Pi-plus-coherent}
\end{equation}
In the above derivation, we use a relation,
\begin{equation}
\sum_{n=0}^{\infty}
|2n\rangle\langle 2n|\alpha-\beta\rangle
=
(1/2)[|\alpha-\beta\rangle +|-(\alpha-\beta)\rangle].
\end{equation}

Similarly, the probability that we obtain $O_{1}=-1$ for the measurement of the initial state $|\alpha\rangle$
at time $t_{1}=0$ is written in the form,
\begin{eqnarray}
P_{1-}
&=&
\langle\alpha|\Pi^{(-)}(\beta)|\alpha\rangle \nonumber \\
&=&
\exp(-|\alpha-\beta|^{2})
\sinh|\alpha-\beta|^{2}.
\end{eqnarray}
Then, the initial state $|\alpha\rangle$ reduces to $w_{1-}(0)/P_{1-}$,
where $w_{1-}(0)$ is given by
\begin{equation}
w_{1-}(0)
=
\Pi^{(-)}(\beta)|\alpha\rangle\langle\alpha|\Pi^{(-)}(\beta),
\label{w1minus0-0}
\end{equation}
\begin{equation}
\Pi^{(-)}(\beta)|\alpha\rangle
=
(1/2)[
|\alpha\rangle
-
\exp(\beta^{*}\alpha-\beta\alpha^{*})|2\beta-\alpha\rangle
].
\label{Pi-minus-coherent}
\end{equation}
In the above derivation, we use a relation,
\begin{equation}
\sum_{n=0}^{\infty}
|2n+1\rangle\langle 2n+1|\alpha-\beta\rangle
=
(1/2)[|\alpha-\beta\rangle -|-(\alpha-\beta)\rangle].
\end{equation}

According to Eq.~(\ref{time-evolution-alpha-beta-0}),
$w_{1\pm}(0)$,
the unnormalized state at time $t_{1}=0$,
evolves into the state $w_{1\pm}(\tau)$ at time $t_{2}=\tau$,
whose explicit form is given by Eq.~(\ref{formula-w1pmtau})
in Appendix~\ref{section-appendix-A}.
Describing the probability that we detect $O_{2}=1$ with the observation of $w_{1\pm}(\tau)$ at time $t_{2}=\tau$ as $p_{1\pm,2+}$,
we obtain it in the form of Eq.~(\ref{p1pm2+}) in Appendix~\ref{section-appendix-A}.
Similarly,
writing down the probability that we have $O_{2}=-1$ with the observation of $w_{1\pm}(\tau)$ at time $t_{2}=\tau$ as $p_{1\pm,2-}$,
we acquire it in the form of Eq.~(\ref{p1pm2-}) in Appendix~\ref{section-appendix-A}.

From Eqs.~(\ref{p1pm2+}) and (\ref{p1pm2-}),
the correlation function $C_{21}$ is given by
\begin{equation}
C_{21}=p_{1+,2+}-p_{1+,2-}-p_{1-,2+}+p_{1-,2-}.
\label{C21-formula-0}
\end{equation}
Here, we regard $C_{21}$ as a function of multiple variables,
\begin{equation}
C_{21}=C(\alpha,\beta,\omega,\Gamma,\tau).
\label{C21-formula-1}
\end{equation}
Then, $C_{31}$ and $C_{32}$ are given by
\begin{equation}
C_{31}=C(\alpha,\beta,\omega,\Gamma,2\tau),
\label{C31-formula-1}
\end{equation}
\begin{equation}
C_{32}=C(\alpha e^{-i\Omega\tau},\beta,\omega,\Gamma,\tau).
\label{C32-formula-1}
\end{equation}
Thus,
from Eqs.~(\ref{C21-formula-0}), (\ref{C21-formula-1}), (\ref{C31-formula-1}), and (\ref{C32-formula-1}),
we obtain $K_{3}$ eventually.

\section{\label{section-LG-equation-for-CAT-state}
The explicit form of the LGI for the system initialized in the cat state}
In this section, preparing the cat state $(|\alpha\rangle +|-\alpha\rangle)$ as the initial state,
we derive a mathematically rigorous form of the LGI.

We assume that the initial state of the system at time $t_{1}=0$ is given by the cat state,
\begin{equation}
\rho(t_{1})=|\psi\rangle\langle\psi|,
\label{initial-state-0}
\end{equation}
\begin{equation}
|\psi\rangle=q(\alpha)^{-1/2}(|\alpha\rangle+|-\alpha\rangle),
\label{initial-state-1}
\end{equation}
\begin{equation}
q(\alpha)=2[1+\exp(-2|\alpha|^{2})],
\label{initial-state-2}
\end{equation}
where $|\alpha\rangle$ and $-|\alpha\rangle$ are coherent states.
It is convenient to use the following notation for calculations carried out hereafter:
\begin{equation}
\rho(t_{1})
=
q(\alpha)^{-1}(K+L+L^{\dagger}+M),
\end{equation}
\begin{eqnarray}
K
&=&
|\alpha\rangle\langle\alpha|, \nonumber \\
L
&=&
|\alpha\rangle\langle -\alpha|, \nonumber \\
M
&=&
|-\alpha\rangle\langle -\alpha|.
\end{eqnarray}

The probability that we obtain $O_{1}=\pm 1$ with the observation of $\rho(t_{1})$ at time $t_{1}=0$ is written down in the form,
\begin{equation}
P_{1\pm}=\mbox{Tr}[\Pi^{(\pm)}(\beta)\rho(t_{1})].
\end{equation}
Then, the state collapses from $\rho(t_{1})$ to $w_{1\pm}(0)/P_{1\pm}$,
where $w_{1\pm}(0)$ is given by
\begin{eqnarray}
w_{1\pm}(0)
&=&
\Pi^{(\pm)}(\beta)\rho(t_{1})\Pi^{(\pm)}(\beta) \nonumber \\
&=&
q(\alpha)^{-1}
\Pi^{(\pm)}(\beta)
(K+L+L^{\dagger}+M)
\Pi^{(\pm)}(\beta).
\label{w1pm-0-definition}
\end{eqnarray}

Moreover, we divide operators $K$, $L$, and $M$ into operators as
\begin{eqnarray}
\Pi^{(\pm)}(\beta)K\Pi^{(\pm)}(\beta)
&=&
(1/4)[K^{(1)}(0)
\pm K^{(2)}(0)
\pm K^{(3)}(0)
+ K^{(4)}(0)], \nonumber \\
\Pi^{(\pm)}(\beta)L\Pi^{(\pm)}(\beta)
&=&
(1/4)[L^{(1)}(0)
\pm L^{(2)}(0)
\pm L^{(3)}(0)
+ L^{(4)}(0)], \nonumber \\
\Pi^{(\pm)}(\beta)M\Pi^{(\pm)}(\beta)
&=&
(1/4)[M^{(1)}(0)
\pm M^{(2)}(0)
\pm M^{(3)}(0)
+ M^{(4)}(0)],
\label{PipKLMPip-definition}
\end{eqnarray}
where $K^{(3)}(0)=K^{(2)}(0)^{\dagger}$ and $M^{(3)}(0)=M^{(2)}(0)^{\dagger}$.
We give explicit forms of
$\{K^{(j)}(0):j=1,2,4\}$,
$\{L^{(j)}(0):j=1,2,3,4\}$, and
$\{M^{(j)}(0):j=1,2,4\}$
in Eqs.~(\ref{K-0-definition}), (\ref{L-0-definition}), and (\ref{M-0-definition}) in Appendix~\ref{section-appendix-B}.
Closed-form expressions of time evolution from time $t_{1}=0$ to time $t_{2}=\tau$ of these operators,
$\{K^{(j)}(\tau):j=1,2,4\}$,
$\{L^{(j)}(\tau):j=1,2,3,4\}$, and
$\{M^{(j)}(\tau):j=1,2,4\}$,
are given by Eqs.~(\ref{K-tau-definition}), (\ref{L-tau-definition}), and (\ref{M-tau-definition})
in Appendix~\ref{section-appendix-B} explicitly.

We consider that the state $w_{1\pm}(0)$ at time $t_{1}=0$ evolves into $w_{1\pm}(\tau)$ at time $t_{2}=\tau$.
Then, we can obtain the state $w_{1\pm}(\tau)$ by replacing
$\{K^{(j)}(0):j=1,2,4\}$,
$\{L^{(j)}(0):j=1,2,3,4\}$, and
$\{M^{(j)}(0):j=1,2,4\}$
in Eqs.~(\ref{w1pm-0-definition}) and (\ref{PipKLMPip-definition})
with
$\{K^{(j)}(\tau):j=1,2,4\}$,
$\{L^{(j)}(\tau):j=1,2,3,4\}$, and
$\{M^{(j)}(\tau):j=1,2,4\}$
given by
Eqs.~(\ref{K-tau-definition}),
(\ref{L-tau-definition}), and
(\ref{M-tau-definition}).
Describing the probability that we detect $O_{2}=1$ with the observation of the state $w_{1\pm}(\tau)$ at time $t_{2}=\tau$
as $p_{1\pm,2+}$,
we obtain it in the form of
Eq.~(\ref{p1pm2p}) in Appendix~\ref{section-appendix-B}.

Letting the probability that we obtain $O_{2}=-1$ with the observation of $w_{1\pm}(\tau)$ at time $t_{2}=\tau$ be equal to $p_{1\pm,2-}$,
we can write down it as
Eq.~(\ref{p1pm2m}) in Appendix~\ref{section-appendix-B}.
Then, we can derive the correlation function $C_{21}$ from
Eq.~(\ref{C21-formula-0}).
We can also obtain $C_{31}$ from Eqs.~(\ref{C21-formula-1}) and (\ref{C31-formula-1}).

Next, we think about the correlation function $C_{32}$.
Here, we remark that we cannot obtain $C_{32}$ from Eq.~(\ref{C32-formula-1}).
The reason why is going to be clarified by Eqs.(\ref{rho-t2-0}) and (\ref{rho-t2-1}).
The initial state at time $t_{1}=0$ is given by Eqs.~(\ref{initial-state-0}), (\ref{initial-state-1}), and (\ref{initial-state-2}).
The initial state $\rho(t_{1})$ evolves into the following state at time $t_{2}=\tau$:
\begin{eqnarray}
\rho(t_{2})
&=&
q(\alpha)^{-1}
\Bigl(
\tilde{K}
+
\exp\{-2|\alpha|^{2}[1-\exp(-2\Gamma\tau)]\}(\tilde{L}+\tilde{L}^{\dagger})
+
\tilde{M}
\Bigr),
\label{rho-t2-0}
\end{eqnarray}
\begin{eqnarray}
\tilde{K}
&=&
|\alpha\exp(-i\Omega\tau)\rangle\langle\alpha\exp(-i\Omega\tau)|, \nonumber \\
\tilde{L}
&=&
|\alpha\exp(-i\Omega\tau)\rangle\langle -\alpha\exp(-i\Omega\tau)|, \nonumber \\
\tilde{M}
&=&
|-\alpha\exp(-i\Omega\tau)\rangle\langle -\alpha\exp(-i\Omega\tau)|.
\label{rho-t2-1}
\end{eqnarray}
Looking at Eqs.~(\ref{rho-t2-0}) and (\ref{rho-t2-1}),
we notice that $\rho(t_{2})$ is different from
$\rho'=|\psi'\rangle\langle\psi'|$
where
$|\psi'\rangle=q(\alpha')^{-1/2}(|\alpha'\rangle+|-\alpha'\rangle)$
and
$\alpha'=\alpha\exp(-i\Omega\tau)$.
Thus, we cannot derive $C_{32}$ from Eq.~(\ref{C32-formula-1}).

The probability that we obtain $O_{2}=\pm 1$ with the observation of $\rho(t_{2})$ at time $t_{2}=\tau$ is given by
\begin{equation}
P_{2\pm}=\mbox{Tr}[\Pi^{(\pm)}(\beta)\rho(t_{2})].
\end{equation}
Then, the state of the system $\rho(t_{2})$ reduces to
$w_{2\pm}(\tau)/P_{2\pm}$,
where $w_{2\pm}(\tau)$ is given by
\begin{eqnarray}
w_{2\pm}(\tau)
&=&
\Pi^{(\pm)}(\beta)\rho(t_{2})\Pi^{(\pm)}(\beta) \nonumber \\
&=&
q(\alpha)^{-1}
\Pi^{(\pm)}(\beta)
\Bigl(
\tilde{K}
+
\exp\{-2|\alpha|^{2}[1-\exp(-2\Gamma\tau)]\}
(\tilde{L}+\tilde{L}^{\dagger})
+
\tilde{M}
\Bigr)
\Pi^{(\pm)}(\beta). \nonumber \\
\label{w-2pm-tau-definition}
\end{eqnarray}

We can divide
$\Pi^{(\pm)}(\beta)\tilde{K}\Pi^{(\pm)}(\beta)$, $\Pi^{(\pm)}(\beta)\tilde{L}\Pi^{(\pm)}(\beta)$, and $\Pi^{(\pm)}(\beta)\tilde{M}\Pi^{(\pm)}(\beta)$
into operators,
\begin{eqnarray}
\Pi^{(\pm)}(\beta)
\tilde{K}
\Pi^{(\pm)}(\beta)
&=&
(1/4)
[
\tilde{K}^{(1)}(0)
\pm
\tilde{K}^{(2)}(0)
\pm
\tilde{K}^{(2)}(0)^{\dagger}
+
\tilde{K}^{(4)}(0)
],
\label{Pipm-tilde-K-Pipm-definition}
\end{eqnarray}
\begin{equation}
\tilde{K}^{(j)}(0)
=
\left. K^{(j)}(0) \right|_{\alpha\rightarrow \alpha\exp(-i\Omega\tau)}
\quad
\mbox{for $j=1,2,3,4$},
\label{tilde-K-0-definition}
\end{equation}
\begin{eqnarray}
\Pi^{(\pm)}(\beta)
\tilde{L}
\Pi^{(\pm)}(\beta)
&=&
(1/4)
[
\tilde{L}^{(1)}(0)
\pm
\tilde{L}^{(2)}(0)
\pm
\tilde{L}^{(2)}(0)^{\dagger}
+
\tilde{L}^{(4)}(0)
],
\label{Pipm-tilde-L-Pipm-definition}
\end{eqnarray}
\begin{equation}
\tilde{L}^{(j)}(0)
=
\left. L^{(j)}(0) \right|_{\alpha\rightarrow \alpha\exp(-i\Omega\tau)}
\quad
\mbox{for $j=1,2,3,4$},
\label{tilde-L-0-definition}
\end{equation}
\begin{eqnarray}
\Pi^{(\pm)}(\beta)
\tilde{M}
\Pi^{(\pm)}(\beta)
&=&
(1/4)
[
\tilde{M}^{(1)}(0)
\pm
\tilde{M}^{(2)}(0)
\pm
\tilde{M}^{(2)}(0)^{\dagger}
+
\tilde{M}^{(4)}(0)
],
\label{Pipm-tilde-M-Pipm-definition}
\end{eqnarray}
\begin{equation}
\tilde{M}^{(j)}(0)
=
\left. M^{(j)}(0) \right|_{\alpha\rightarrow \alpha\exp(-i\Omega\tau)}
\quad
\mbox{for $j=1,2,3,4$}.
\label{tilde-M-0-definition}
\end{equation}
For example, Eq.~(\ref{tilde-K-0-definition}) implies the following.
The operator $\tilde{K}^{(j)}(0)$ is equal to the operator $K^{(j)}(0)$ with the replacement of $\alpha$ by $\alpha\exp(-i\Omega\tau)$
in Eq.~(\ref{K-0-definition}).

Referring to Eqs.~(\ref{K-tau-definition}),
(\ref{L-tau-definition}), and
(\ref{M-tau-definition}),
closed-form expressions of time evolution from time $t_{2}=\tau$ to time $t_{3}=2\tau$ of
$\{\tilde{K}^{(j)}(0):j=1,2,3,4\}$,
$\{\tilde{L}^{(j)}(0):j=1,2,3,4\}$, and
$\{\tilde{M}^{(j)}(0):j=1,2,3,4\}$
are given by
\begin{eqnarray}
\tilde{K}^{(j)}(\tau)
&=&
\left. K^{(j)}(\tau) \right|_{\alpha\rightarrow \alpha\exp(-i\Omega\tau)}, \nonumber \\
\tilde{L}^{(j)}(\tau)
&=&
\left. L^{(j)}(\tau) \right|_{\alpha\rightarrow \alpha\exp(-i\Omega\tau)}, \nonumber \\
\tilde{M}^{(j)}(\tau)
&=&
\left. M^{(j)}(\tau) \right|_{\alpha\rightarrow \alpha\exp(-i\Omega\tau)}
\quad
\mbox{for $j=1,2,3,4$}.
\label{tilde-KLM-tau-definition}
\end{eqnarray}
Thus,
the state $w_{2\pm}(2\tau)$,
that is to say,
the time evolution from time $t_{2}=\tau$ to time $t_{3}=2\tau$ of the state $w_{2\pm}(\tau)$,
is obtained by replacing
$\{\tilde{K}^{(j)}(0):j=1,2,3,4\}$,
$\{\tilde{L}^{(j)}(0):j=1,2,3,4\}$, and
$\{\tilde{M}^{(j)}(0):j=1,2,3,4\}$
with Eq.~(\ref{tilde-KLM-tau-definition})
in Eqs.~(\ref{w-2pm-tau-definition}),
(\ref{Pipm-tilde-K-Pipm-definition}),
(\ref{tilde-K-0-definition}),
(\ref{Pipm-tilde-L-Pipm-definition}),
(\ref{tilde-L-0-definition}),
(\ref{Pipm-tilde-M-Pipm-definition}), and
(\ref{tilde-M-0-definition}).

We describe the probability that we obtain $O_{3}=1$ with the observation of $w_{2\pm}(2\tau)$ at time $t_{3}=2\tau$ as $p_{2\pm,3+}$.
Similarly, we let $p_{2\pm,3-}$ represent the probability that we obtain $O_{3}=-1$ with the observation of $w_{2\pm}(2\tau)$ at time $t_{3}=2\tau$.
Then, they are given by
Eqs.~(\ref{p2pm2p-formula}) and (\ref{p2pm2m-formula})
in Appendix~\ref{section-appendix-B}.

Thus, the correlation function $C_{32}$ is given in the form,
\begin{equation}
C_{32}=p_{2+,3+}-p_{2+,3-}-p_{2-,3+}+p_{2-,3-}.
\label{C32-formula-0}
\end{equation}
Because we have just derived $C_{21}$, $C_{31}$, and $C_{32}$,
finally we obtain $K_{3}$.

Here,
we point out that $K_{3}$ has the following symmetry about the variable $\beta$.
The value of $K_{3}$ is invariant under a transformation $\beta\rightarrow -\beta$.
This fact is understood from relations,
\begin{eqnarray}
\mbox{Tr}[K^{(1)}(\tau)\Pi^{(+)}(-\beta)]
&=&
\mbox{Tr}[M^{(1)}(\tau)\Pi^{(+)}(\beta)], \nonumber \\
\mbox{Tr}[K^{(2)}(\tau)\Pi^{(+)}(-\beta)]
&=&
\mbox{Tr}[M^{(2)}(\tau)\Pi^{(+)}(\beta)], \nonumber \\
\mbox{Tr}[K^{(4)}(\tau)\Pi^{(+)}(-\beta)]
&=&
\mbox{Tr}[M^{(4)}(\tau)\Pi^{(+)}(\beta)], \nonumber \\
\mbox{Tr}[L^{(1)}(\tau)\Pi^{(+)}(-\beta)]
&=&
\mbox{Tr}[L^{(1)}(\tau)\Pi^{(+)}(\beta)]^{*}, \nonumber \\
\mbox{Tr}[L^{(2)}(\tau)\Pi^{(+)}(-\beta)]
&=&
\mbox{Tr}[L^{(3)}(\tau)\Pi^{(+)}(\beta)]^{*}, \nonumber \\
\mbox{Tr}[L^{(3)}(\tau)\Pi^{(+)}(-\beta)]
&=&
\mbox{Tr}[L^{(2)}(\tau)\Pi^{(+)}(\beta)]^{*}, \nonumber \\
\mbox{Tr}[L^{(4)}(\tau)\Pi^{(+)}(-\beta)]
&=&
\mbox{Tr}[L^{(4)}(\tau)\Pi^{(+)}(\beta)]^{*}.
\end{eqnarray}

Hence,
putting $\beta=r\exp(i\theta)$ and examining a value of $K_{3}$ with variation of $\beta$,
we only have to try $\theta$ in a range of $0\leq\theta\leq\pi$.
In other words,
$K_{3}$ is a periodic function of $\theta$ and its period is equal to $\pi$.

\section{\label{section-initial-coherent-state-numerical-analyses}
Numerical analyses of the LGI and its optimization with initially putting the system in a coherent state $|\alpha\rangle$}
In the current section,
preparing the system initially in a coherent state $|\alpha\rangle$
for the LGI,
we compute $K_{3}$ numerically and examine its properties.
Moreover, we go into the optimization problem of the displaced parity operators.
In order to let discussion of the current section be simple,
we assume
\begin{equation}
\alpha=1/2,
\quad
\omega=1.
\end{equation}

Because we set $\omega=1$, that is we let the angular velocity be equal to unity,
we can adopt the following system of units.
Because of $\omega=1$, we can rewrite a dimensionless quantity $\omega\tau$ as $\tau$.
Thus,
we can regard the time variable $\tau$ as dimensionless.
Similarly,
we can rewrite a dimensionless quantity $\Gamma/\omega$ as $\Gamma$,
so that we can consider $\Gamma$ to be dimensionless, as well.
Hereafter,
we use this system of units.

Here, we define a simple notation as follows.
Fixing the variables $\alpha$ and $\omega$ at $\alpha=1/2$ and $\omega=1$ respectively,
we describe $K_{3}$ as a function of $\Gamma$, $\beta=re^{i\theta}$, and $\tau$,
\begin{equation}
K_{3}(\Gamma;\theta,r,\tau).
\end{equation}

\begin{figure}
\begin{center}
\includegraphics{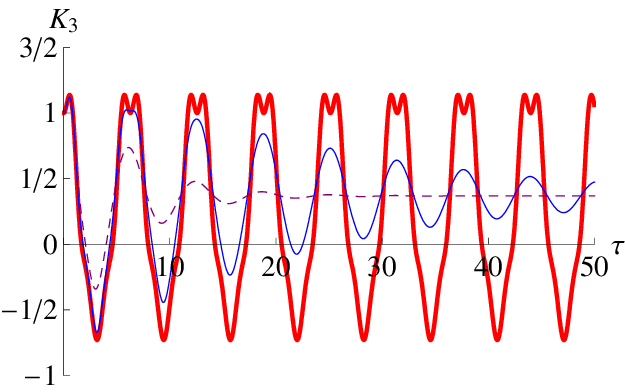}
\end{center}
\caption{Graphs of $K_{3}(\Gamma;0,1/2,\tau)$ as a function of $\tau$ with
$\beta=1/2$, that is $\theta=0$ and $r=1/2$,
where the system is initialized in the coherent state.
The thick solid red, thin solid blue, and thin dashed purple curves represent plots of $\Gamma=0$, $0.05$, and $0.2$, respectively.
The thick solid red curve has a period $2\pi$.
We notice that amplitudes of the curves shrink as $\Gamma$ becomes larger.}
\label{Figure01}
\end{figure}

In Fig.~\ref{Figure01}, we plot $K_{3}(\Gamma;0,1/2,\tau)$ as a function of $\tau$ with $\beta=1/2$, that is $\theta=0$ and $r=1/2$.
The thick solid red, thin solid blue, and thin dashed purple curves represent graphs with $\Gamma=0$, $0.05$, and $0.2$, respectively.
The graph of the thick solid red curve has a period $2\pi$.
As $\Gamma$ becomes larger, amplitudes of the curves shrink.

We can confirm numerically that the curve of $\Gamma=0.2$ converges to $0.367{\,}879...$ as $\tau$ increases.
We can also validate this fact in the manner of analytical mathematics.
Fixing $\Gamma$ at a finite positive value $\Gamma_{0}(>0)$ and letting $\tau$ approach infinity, $\tau\to\infty$,
we obtain
\begin{eqnarray}
\lim_{\tau\rightarrow\infty}K_{3}(\Gamma_{0};0,1/2,\tau)
&=&
1/e \nonumber \\
&\simeq&
0.367{\,}879....
\end{eqnarray}
From the above result,
we understand that $K_{3}$ converges to $1/e$ for $\Gamma_{0}>0$ and $\tau\rightarrow\infty$.

In general,
we can show the following.
$\forall\alpha$, $\forall\omega$, $\forall\Gamma(>0)$, and $\forall\beta=re^{i\theta}$, we obtain
\begin{equation}
\lim_{\tau\rightarrow\infty}K_{3}
=
\exp(-4r^{2}).
\label{K3-limit-tau-infty-coherent}
\end{equation}
We can explain that Eq.~(\ref{K3-limit-tau-infty-coherent}) does not depend on $\omega$ from the following consideration.
According to Eq.~(\ref{time-evolution-alpha-beta-0}),
the angular velocity $\omega$ appears in the explicit expression of $K_{3}$
as the form $\exp(-i\Omega\tau)$ with $\Omega=\omega-i\Gamma$.
Thus, setting $\Gamma>0$,
we obtain $\exp(-i\Omega\tau)\rightarrow 0$ for $\tau\rightarrow\infty$ and dependency of $\omega$ disappears.

From Eq.~(\ref{K3-limit-tau-infty-coherent}),
we become aware that $K_{3}\leq 1$ for $\tau\rightarrow\infty$.
Thus, we understand that the violation of the LGI vanishes as $\tau$ approaches infinity.

\begin{figure}
\begin{center}
\includegraphics{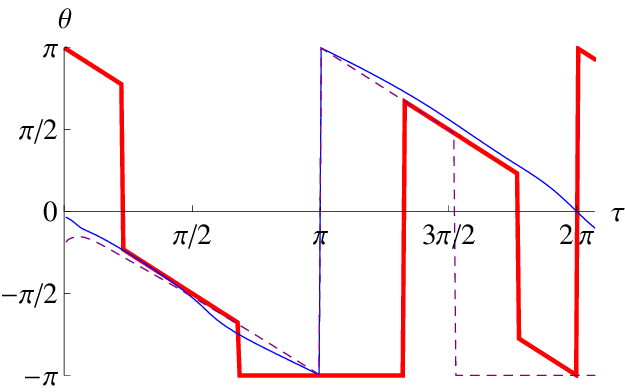}
\end{center}
\caption{Graphs of the optimum $\theta$ that maximizes $K_{3}$ for each $\tau(>0)$
as a function of $\tau$
with letting the system be initialized in the coherent state.
The thick solid red, thin solid blue, and thin dashed purple curves represent plots of $\Gamma=0$, $0.1$, and $1$, respectively.
For all the three curves,
we can find discontinuity points.}
\label{Figure02}
\end{figure}

\begin{figure}
\begin{center}
\includegraphics{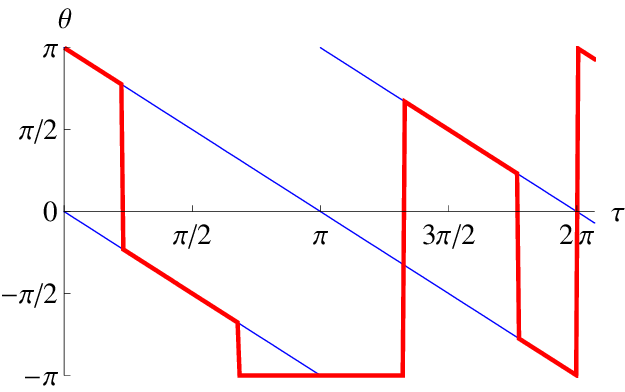}
\end{center}
\caption{A thick red curve represents a plot of the optimum $\theta$ that maximizes $K_{3}$ for each $\tau(>0)$
as a function of $\tau$ with $\Gamma=0$,
where the system is prepared initially in the coherent state.
Three parallel thin blue lines represent graphs of
$\theta=2\pi-\tau$, $\theta=\pi-\tau$, and $\theta=-\tau$.
The optimized $\theta$ moves and jumps on the lines of
$\theta=\pi-\tau$,
$\theta=-\tau$,
$\theta=-\pi$,
$\theta=2\pi-\tau$, and
$\theta=\pi-\tau$
in order as $\tau$ increases from zero to $6.5$.
Because $K_{3}$ is periodic about $\theta$ and its period is given by $2\pi$,
the lines
$\theta=2\pi-\tau$ and $\theta=-\tau$ are essentially equivalent to each other.}
\label{Figure03}
\end{figure}

\begin{figure}
\begin{center}
\includegraphics{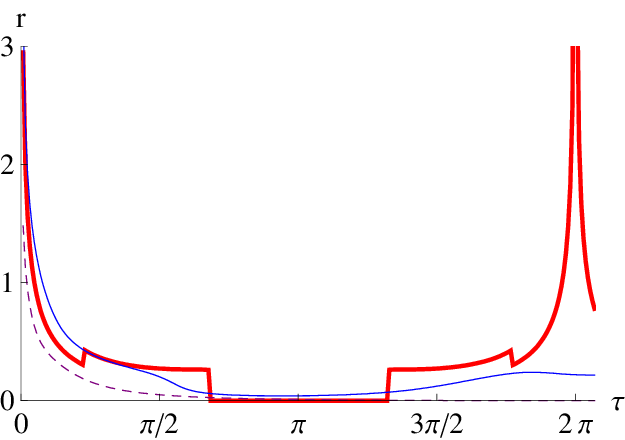}
\end{center}
\caption{Graphs of the optimum $r$ that maximizes $K_{3}$ for each $\tau(>0)$
as a function of $\tau$,
where the system is initialized in the coherent state.
The thick solid red, thin solid blue, and thin dashed purple curves represent plots of $\Gamma=0$, $0.1$, and $1$, respectively.
All the three curves seemingly diverge to infinity at $\tau=0$.
Moreover, the thick red curve of $\Gamma=0$ apparently diverges to infinity at $\tau=2\pi$.}
\label{Figure04}
\end{figure}

\begin{figure}
\begin{center}
\includegraphics{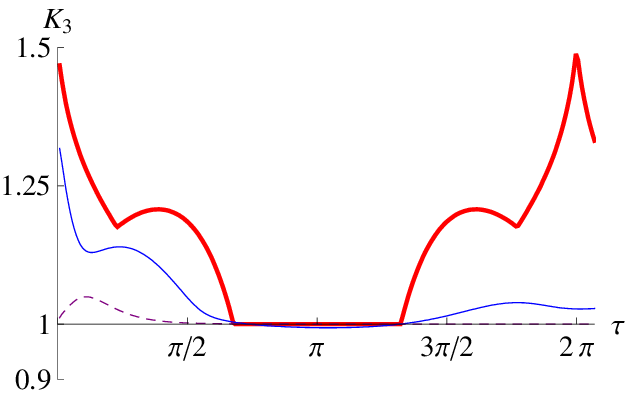}
\end{center}
\caption{Graphs of the maximized $K_{3}$ with adjusting $\theta$ and $r$ for each $\tau(>0)$
as a function of $\tau$
with letting the system be prepared initially in the coherent state.
The thick solid red, thin solid blue, and thin dashed purple curves represent plots of $\Gamma=0$, $0.1$, and $1$, respectively.
As $\Gamma$ increases, amplitudes of the graphs are suppressed.}
\label{Figure05}
\end{figure}

Next,
we consider the following optimization problem.
At given arbitrary $\tau(>0)$,
we look for values of $\theta$ and $r$ which maximize $K_{3}$.
In Figs.~\ref{Figure02} and \ref{Figure04},
we plot the optimum $\theta$ and $r$ that maximize $K_{3}$ for each $\tau(>0)$ as functions of $\tau$.
In Fig.~\ref{Figure05},
we draw a curve of the maximized $K_{3}$ versus $\tau$.
In Figs.~\ref{Figure02}, \ref{Figure04}, and \ref{Figure05},
the thick solid red, thin solid blue, and thin dashed purple curves represent plots for $\Gamma=0$, $0.1$, and $1$, respectively.
Moreover, in Fig.~\ref{Figure03}, we plot the optimum $\theta$ that maximizes $K_{3}$ for each $\tau(>0)$ and $\Gamma=0$ as a function of $\tau$ with a thick red curve
and draw $\theta=2\pi-\tau$, $\theta=\pi-\tau$, and $\theta=-\tau$ with thin blue lines.
Drawing graphs in Figs.~\ref{Figure02}, \ref{Figure03}, \ref{Figure04}, and \ref{Figure05},
we divide a range $0< \tau\leq 6.5$ into equal spaces as $\tau=n\Delta\tau$, $\Delta\tau=0.025$, and $n=1,2,...,260$
and optimize $\theta$ and $r$ at each time $\tau=n\Delta\tau$.

Here, we concentrate on Fig.~\ref{Figure03}.
The optimum $\theta$ that maximizes $K_{3}$ moves and jumps on three parallel lines,
$\theta=2\pi-\tau$, $\theta=\pi-\tau$, and $\theta=-\tau$.
In the following paragraphs,
we
analyse this fact in detail.

Carrying out slightly tough calculations, we can show a relation,
\begin{eqnarray}
\left.\frac{\partial}{\partial\theta}K_{3}(0;\theta,r,\tau)\right|_{\theta=\pi-\tau}
&=&
\left.\frac{\partial}{\partial\theta}K_{3}(0;\theta,r,\tau)\right|_{\theta=-\tau} \nonumber \\
&=&
0.
\end{eqnarray}
Because $K_{3}(0;\theta,r,\tau)$ is a periodic function about $\theta$ and its period is equal to $2\pi$,
it is obvious that
$\left.(\partial/\partial\theta)K_{3}(0;\theta,r,\tau)\right|_{\theta=2\pi-\tau}=0$ holds.
Hence, for three cases where
$\theta=2\pi-\tau$, $\theta=\pi-\tau$, and $\theta=-\tau$,
we can expect that $K_{3}(0;\theta,r,\tau)$ takes extreme values.
Now,
we examine this expectation concretely below.

\begin{figure}
\begin{center}
\includegraphics{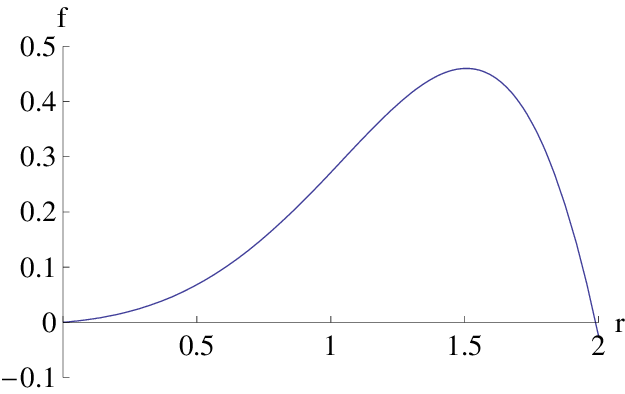}
\end{center}
\caption{A graph of $f(r)=\left.(\partial/\partial r)K_{3}(0;\pi-\tau,r,\tau)\right|_{\tau=0.05}$ versus $r$.
For $r=0$ and $r=1.989{\,}01...$,
$f(r)=0$ holds.}
\label{Figure06}
\end{figure}

Figure~\ref{Figure06} shows a plot of $f(r)$ versus $r$,
where $f(r)$ is given by
\begin{equation}
f(r)=\left.\frac{\partial}{\partial r}K_{3}(0;\pi-\tau,r,\tau)\right|_{\tau=0.05}.
\end{equation}
Looking at Fig.~\ref{Figure06},
we notice that $f(r)=0$ holds at points,
$r=0$ and $r=1.989{\,}01...$.
For $\tau=0.05$,
we can maximize $K_{3}$ at $r=1.989{\,}01...$.

Because of these circumstances,
the optimum $\theta$ that maximizes $K_{3}$ walks and jumps around three lines,
$\theta=2\pi-\tau$, $\theta=\pi-\tau$, and $\theta=-\tau$.
In Fig.~\ref{Figure03},
in a range of $2.150\leq\tau\leq 4.150$,
$\theta=-\pi$ holds.
In this range with $\Gamma=0$,
we can confirm that $r=0$ holds by looking at Fig.~\ref{Figure04}.
Thus, in the range of $2.150\leq\tau\leq 4.150$,
we can consider that the value of $\theta$ is meaningless.

Around a neighbourhood of $\tau=0$,
the function of the maximized $K_{3}$ exhibits strong singularity.
According to Eq.~(\ref{definition-Leggett-Garg-inequality-0}),
the definition of $K_{3}$,
we can naively suppose
\begin{equation}
K_{3}(0;\theta,r,0)=1.
\end{equation}
However,
inspecting Fig.~\ref{Figure05},
we recognize that the maximized $K_{3}$ approaches $1.5$ in the limit $\tau\rightarrow +0$.
By numerical calculations,
we can verify that
the maximized $K_{3}$ attains $1.498{\,}48...$ for $\tau=0.001$.
From these careful looks,
we grasp that it is very difficult to estimate the maximum value of $K_{3}$ in the limit $\tau\to +0$.
In fact,
Fig.~\ref{Figure04} shows that the optimized $r$ apparently diverges to infinity,
$r\rightarrow\infty$,
as $\tau$ approaches zero, $\tau\rightarrow +0$.
This fact makes the problem be very intractable.
In the following paragraphs,
we
consider this problem carefully.

\begin{figure}
\begin{center}
\includegraphics{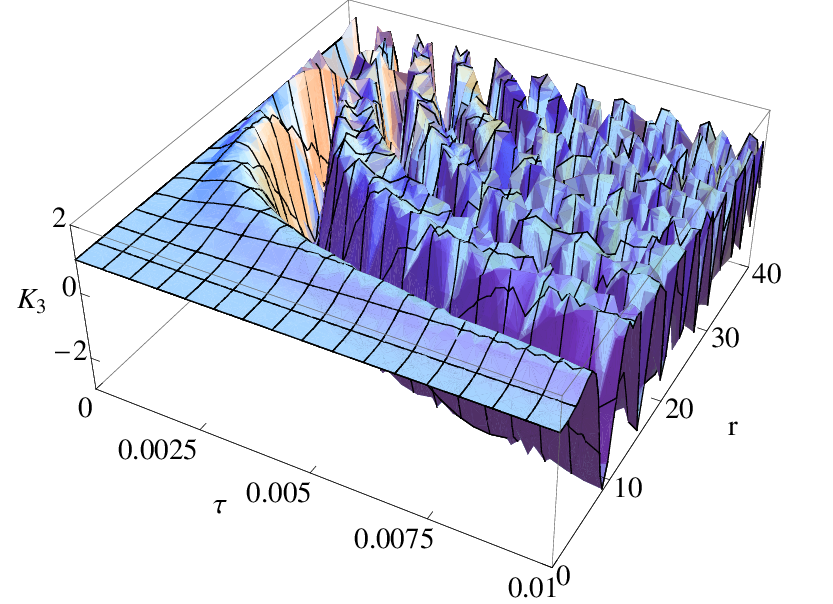}
\end{center}
\caption{A plot of $g(r,\tau)$ given by Eq.~(\ref{g_function})
on a two-dimensional plane of $(\tau,r)$.
We can find strong singularity in the limits, $\tau\rightarrow +0$ and $r\rightarrow\infty$.}
\label{Figure07}
\end{figure}

Paying our attention to Fig.~\ref{Figure03},
we can recognize that the optimum $\theta$ that maximizes $K_{3}$ is given by $\theta=\pi-\tau$.
Thus, we consider a function,
\begin{eqnarray}
g(r,\tau)
&=&
K_{3}(0;\pi-\tau,r,\tau) \nonumber \\
&=&
2\exp[4r^{2}(-1+\cos\tau)]\cos[2r(1+2r)\sin\tau] \nonumber \\
&&
-\exp(-8r^{2}\sin^{2}\tau)\cos[4r(1+2r\cos\tau)\sin\tau].
\label{g_function}
\end{eqnarray}
Figure~\ref{Figure07} shows a graph of $g(r,\tau)$ plotted in a
two-dimensional plane of $(\tau,r)$.
In Fig.~\ref{Figure07},
we can find strong singularity in the limits, $\tau\rightarrow +0$ and $r\rightarrow\infty$,
and we cannot estimate the maximum value of $g(r,\tau)$ under these conditions.

Here, we apply the following approximations to $g(r,\tau)$.
Taking the limit $\tau\rightarrow +0$,
we set $\sin\tau\simeq \tau$ and $\cos\tau\simeq 1-\tau^{2}/2$.
Moreover, we let $r\gg 1$.
Then, we obtain
\begin{equation}
g(r,\tau)
\simeq
2\exp(-2r^{2}\tau^{2})\cos(4r^{2}\tau)-\exp(-8r^{2}\tau^{2})\cos(8r^{2}\tau).
\label{g_approximation}
\end{equation}
Because $g(r,\tau)$ given by Eq.~(\ref{g_approximation}) is very unmanageable,
it is difficult to compute its maximum value under $\tau\rightarrow +0$ and $r\rightarrow\infty$.
Thus, we utilize the following special technique for obtaining the limit of $g(r,\tau)$.

Here, we put $x=r^{2}\tau$ and assume that $x$ takes a finite value.
Then, we can rewrite Eq.~(\ref{g_approximation}) as follows:
\begin{equation}
g(\tau,x)
=
2\exp(-2x\tau)\cos 4x -\exp(-8x\tau)\cos 8x.
\end{equation}
For the above equation,
first we fix $x$ at a finite value and second we let $\tau$ approach zero, that is $\tau\to +0$.
As a result of these operations,
we obtain the following function as the limit of $g(\tau,x)$:
\begin{equation}
g(x)
=
2 \cos 4x-\cos 8x.
\label{Leggett-Garg-tau-zero-limit-coherent}
\end{equation}
The smallest positive value of $x$ that maximizes $g(x)$ is given by $x=\pi/12$
and the maximum of $g(x)$ is equal to $3/2$.
Thus,
we can suppose that the maximum value of $K_{3}(0;\pi-\tau,r,\tau)$ converges to $3/2$ under $\tau\to +0$ and $r\to\infty$.

Here,
we are confronted with a problem whether or not $r^{2}\tau=\pi/12$ actually holds for $r$ and $\tau$
that maximize $K_{3}(0;\pi-\tau,r,\tau)$.
We have confirmed that $r^{2}\tau\simeq\pi/12$ holds around $\tau\simeq 0$ for $\Gamma=0$ by numerical calculations.
To be more precise,
we have obtained that $r^{2}\tau=\mbox{Const.}(\pi/12)$ with $0.9068\leq\mbox{Const.}\leq 0.9683$ for $\tau=0.001\times n$ and $n=1,2,...,8$.
Moreover, we have verified that $\mbox{Const.}$ increases and gets closer to unity as $\tau$ approaches zero.
Thus, this evidence suggests that the maximum $K_{3}$ converges to $3/2$ for $\tau\to +0$.

In Refs.~\cite{Emary2014,Friedenberger2017},
the following results were shown.
We consider a two-level atom which does not interact with an environment and has independent time evolution.
We assume that its Hamiltonian is given by $H=(\omega_{0}/2)\sigma_{z}$.
Then, we obtain the time-symmetrized correlation function,
\begin{eqnarray}
C(\tau)
&=&
\langle\{\sigma_{x}(t),\sigma_{x}(t+\tau)\}\rangle/2 \nonumber \\
&=&
\cos(\omega_{0}\tau).
\end{eqnarray}
Thus, $K_{3}$ of the LGI of equally spaced measurements with separation $\tau$ is given by
\begin{eqnarray}
K_{3}
&=&2C(\tau)-C(2\tau) \nonumber \\
&=&
2\cos(\omega_{0}\tau)-\cos(2\omega_{0}\tau).
\label{Leggett-Garg-two-level-system}
\end{eqnarray}
This equation is similar to Eq.~(\ref{Leggett-Garg-tau-zero-limit-coherent})
and they correspond to each other with putting $\omega_{0}\tau=4x$.
The reason why we can find such a resemblance in Eqs.~(\ref{Leggett-Garg-tau-zero-limit-coherent}) and (\ref{Leggett-Garg-two-level-system})
is as follows.

If we fix $\alpha$ at a finite complex value and let $\beta$ diverge to infinity, $\beta=re^{i\theta}$ and $r\rightarrow\infty$,
we obtain
\begin{eqnarray}
\langle\alpha|2\beta\rangle
&=&
\exp[-\frac{1}{2}(|\alpha|^{2}+4|\beta|^{2})+2\alpha^{*}\beta] \nonumber \\
&\rightarrow &
0,
\label{approximation-orthogonality-alpha-2beta}
\end{eqnarray}
so that we can consider that $|\alpha\rangle$ and $|2\beta\rangle$ are approximately orthogonal to each other.
Because of Eqs.~(\ref{Pi-plus-coherent}) and (\ref{Pi-minus-coherent}),
putting $|\beta|\rightarrow\infty$,
we obtain
\begin{eqnarray}
\Pi^{(\pm)}(\beta)|\alpha\rangle
&=&
\frac{1}{2}[
|\alpha\rangle
\pm
\exp(\beta^{*}\alpha-\beta\alpha^{*})|2\beta-\alpha\rangle
] \nonumber \\
&\rightarrow &
\frac{1}{2}(|\alpha\rangle
\pm e^{i\varphi}|2\beta\rangle),
\label{Pi-pm-beta-alpha-limit}
\end{eqnarray}
where $\alpha=se^{i\mu}$ and $\varphi=2irs\sin(\mu-\theta)$.

These facts imply that $\Pi^{(\pm)}(\beta)$ are equivalent to a projection measurement with $\sigma_{x}$
if we regard the system of interest as a two-level system
$\{|\alpha\rangle,e^{i\varphi}|2\beta\rangle\}$,
that is
$|0\rangle\equiv |\alpha\rangle$ and $|1\rangle\equiv e^{i\varphi}|2\beta\rangle$.
Defining the Hamiltonian of the system as $H=\omega a^{\dagger}a$ with $\omega=1$,
we evaluate energies of the two states
$|\alpha\rangle$ and $e^{i\varphi}|2\beta\rangle$ roughly,
\begin{equation}
\langle\alpha|H|\alpha\rangle
=
|\alpha|^{2},
\end{equation}
\begin{equation}
\langle 2\beta|e^{-i\varphi}He^{i\varphi}|2\beta\rangle
=
4r^{2}.
\end{equation}
Because of $|\alpha|^{2}\ll 4r^{2}$,
we can rewrite the Hamiltonian as
\begin{equation}
H=\frac{1}{2}(4r^{2})\sigma_{z}.
\end{equation}
Here, setting $\omega_{0}=4r^{2}$,
we can obtain $K_{3}$ of the LGI for this system as
\begin{equation}
K_{3}
=
2\cos(4r^{2}\tau)-\cos(8r^{2}\tau).
\label{K3-two-level-system}
\end{equation}
If we substitute $r^{2}\tau=x$ into Eq.~(\ref{K3-two-level-system}),
we reach Eq.~(\ref{Leggett-Garg-tau-zero-limit-coherent}).

During the above discussion,
first we take the limit $\tau\to +0$,
and second we let $r$ approach infinity as $r\to\infty$.
Here, we consider these two processes in reverse.
First of all, we choose $\Pi^{(\pm)}(\beta)$ for the projection measurement of the system.
Although $\Pi^{(\pm)}(\beta)|\alpha\rangle$ are orthogonal to each other,
they are not normalized.

To maximize $K_{3}$,
we had better make $\Pi^{(\pm)}(\beta)|\alpha\rangle$ be normalized and orthogonal to each other.
Thus, because of Eqs.~(\ref{approximation-orthogonality-alpha-2beta}) and (\ref{Pi-pm-beta-alpha-limit}),
we need to let $|\beta|$ diverge to infinity as $|\beta|\rightarrow\infty$.
Then, we can regard the system of interest as a two-level system and its $K_{3}$ is given by Eq.~(\ref{K3-two-level-system}).
In Eq.~(\ref{K3-two-level-system}),
in order to maximize $K_{3}$,
we have to let a relation $r^{2}\tau=(\pi/12)+(n/2)\pi$ holds for $n=0,1,2,...$.
Here, we choose the smallest value as $r^{2}\tau=\pi/12$ for determining it uniquely.
Then,
due to $r=|\beta|\rightarrow\infty$,
we come to a conclusion $\tau\rightarrow +0$.
Therefore, we have to employ the two processes of approaching the limits,
$r\rightarrow\infty$ and $\tau\rightarrow +0$,
and the singularity emerges.

In the above arguments,
the limit $r\to\infty$ seems to cause the limit $\tau\to +0$.
However,
from a viewpoint of a practical procedure of physics,
we have to first decide the time $\tau$ of the measurement and second determine the parameter $r$ for the projection operators,
so that we cannot help feeling that the order of taking limits are not normal.
Here, we calm ourselves and follow the discussion in a different manner.
If we take the limit $r\to\infty$,
the optimum $\tau$ that maximizes $K_{3}$ is uniquely determined as $\tau\to +0$
because of $r^{2}\tau=\pi/12$.
Hence,
thinking about the optimization problem,
we have to admit that the two operations, taking the limits as $r\to\infty$ and $\tau\to +0$,
are commutes with each other.

\section{\label{section-initial-CAT-state-numerical-analyses}
Numerical analyses of the LGI and its optimization for the system initially prepared in the cat state $(|\alpha\rangle+|-\alpha\rangle)$}
In the present section,
letting the initial state be given by the cat state,
we examine $K_{3}$ of the LGI numerically.
Furthermore,
we consider the optimization problem of the displaced parity operators.
In a similar fashion to Sect.~\ref{section-initial-coherent-state-numerical-analyses},
to make discussion be simple,
we put $\alpha=1/2$ and $\omega=1$.
Moreover,
we use the notation $K_{3}(\Gamma;\theta,r,\tau)$.

\begin{figure}
\begin{center}
\includegraphics{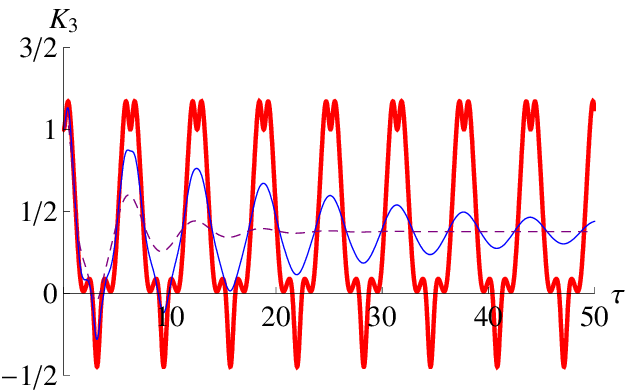}
\end{center}
\caption{Graphs of $K_{3}(\Gamma;0,1/2,\tau)$ as a function of $\tau$ with letting the system be given by the cat state initially and putting $\beta=1/2$,
that is $\theta=0$ and $r=1/2$.
The thick solid red, thin solid blue, and thin dashed purple curves represent plots of $\Gamma=0$, $0.05$, and $0.2$, respectively.
The thick solid red plot is periodic about $\tau$ and its period is equal to $2\pi$.
As $\Gamma$ becomes larger,
amplitudes of the curves decrease.}
\label{Figure08}
\end{figure}

In Fig.~\ref{Figure08},
we draw graphs of $K_{3}(\Gamma;0,1/2,\tau)$ as a function of $\tau$ with $\beta=1/2$,
that is $\theta=0$ and $r=1/2$.
The thick solid red, thin solid blue, and thin dashed purple curves represent plots of $\Gamma=0$, $0.05$, and $0.2$, respectively.
The thick solid red graph is periodic about $\tau$ and its period is given by $2\pi$.
As $\Gamma$ increases, amplitudes of the curves shrink.

We can numerically verify that the curve of $\Gamma=0.2$ converges to $0.376{\,}742...$
as $\tau$ becomes larger.
This fact can be confirmed in the manner of mathematical analysis.
Fixing $\Gamma$ at a finite value $\Gamma_{0}(>)$
and taking the limit $\tau\to\infty$,
we obtain
\begin{eqnarray}
\lim_{\tau\to\infty}
K_{3}(\Gamma_{0};0,1/2,\tau)
&=&
\frac{1+2\sqrt{e}+5e}{4e^{3/2}(1+\sqrt{e})} \nonumber \\
&\simeq&
0.376{\,}472....
\label{limit-K3-cat-state}
\end{eqnarray}
From Eq.~(\ref{limit-K3-cat-state}),
we understand that $K_{3}$ converges to the above value for $\Gamma_{0}>0$ and $\tau\to\infty$.

In general,
$\forall\alpha=se^{i\mu}$, $\forall\omega$, $\forall\Gamma(>0)$, and $\forall\beta=re^{i\theta}$,
we obtain
\begin{eqnarray}
\lim_{\tau\rightarrow\infty}
K_{3}
&=&
\frac{\exp\{-2r[2r+s\cos(\theta-\mu)]\}}{4[1+\exp(2s^{2})]} \nonumber \\
&&
\times
\Biggl\{
\exp[2rs\cos(\theta-\mu)][3+\exp(2r^{2})+4\exp(2s^{2})] \nonumber \\
&&
+
[1-\exp(2r^{2})]\cos[2rs\sin(\theta-\mu)]
\Biggr\}.
\end{eqnarray}

\begin{figure}
\begin{center}
\includegraphics{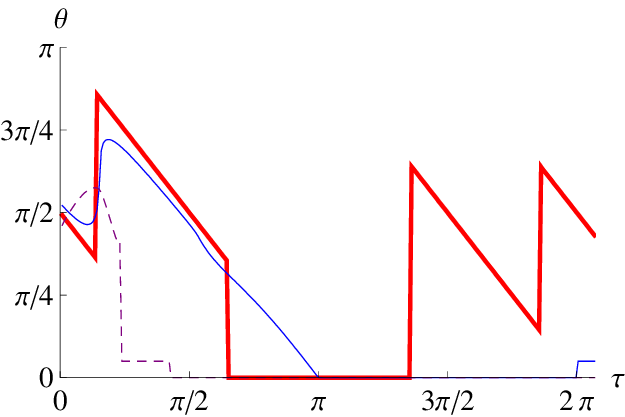}
\end{center}
\caption{Graphs of the optimum $\theta$ that maximizes $K_{3}$ for each $\tau(>0)$
as a function of $\tau$ with letting the system be given by the cat state initially.
The thick solid red, thin solid blue, and thin dashed purple curves represent plots of $\Gamma=0$, $0.1$, and $1$, respectively.
There are discontinuity points in all the three curves.}
\label{Figure09}
\end{figure}

\begin{figure}
\begin{center}
\includegraphics{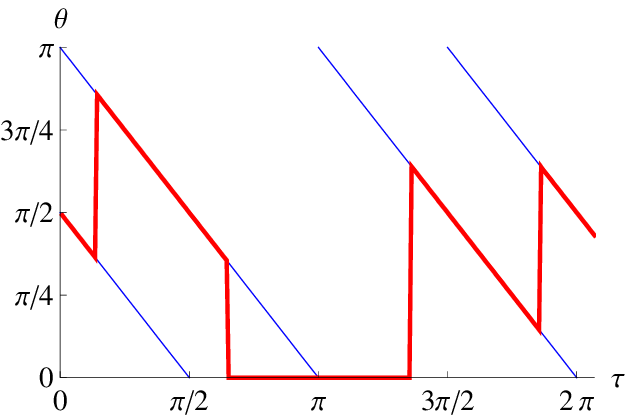}
\end{center}
\caption{A thick red curve represents the optimum $\theta$ that maximizes $K_{3}$
for each $\tau(>0)$
with $\Gamma=0$ as a function of $\tau$,
where the system is initialized in the cat state.
Four parallel thin blue lines represent
$\theta=(5\pi/2)-\tau$, $\theta=2\pi-\tau$, $\theta=\pi-\tau$, and $\theta=(\pi/2)-\tau$.
The optimum $\theta$ moves and jumps on the lines,
$\theta=(\pi/2)-\tau$,
$\theta=\pi-\tau$,
$\theta=-\pi$,
$\theta=2\pi-\tau$, and
$\theta=(5\pi/2)-\tau$,
in order as $\tau$ increases.
As explained in Sect.~\ref{section-LG-equation-for-CAT-state},
$K_{3}$ is a periodic function about $\theta$ and its period is given by $\pi$,
so that
$\theta=2\pi-\tau$ and $\theta=\pi-\tau$
are essentially equivalent to each other,
and so do
$\theta=(5\pi/2)-\tau$ and $\theta=(\pi/2)-\tau$.}
\label{Figure10}
\end{figure}

\begin{figure}
\begin{center}
\includegraphics{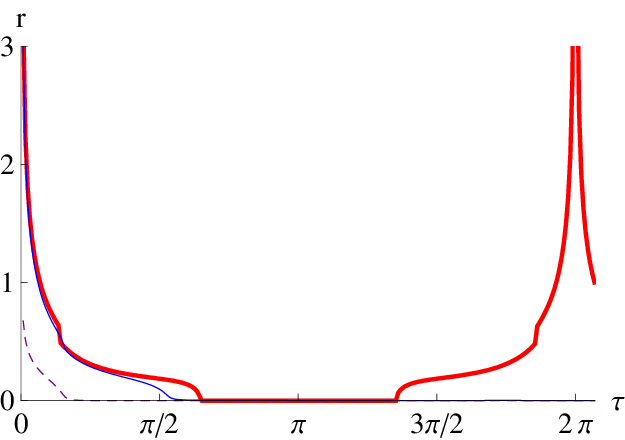}
\end{center}
\caption{Plots of the optimum $r$ that maximizes $K_{3}$ for each $\tau(>0)$
as a function of $\tau$
with preparing the system initially in the cat state.
The thick solid red, thin solid blue, and thin dashed purple curves represent graphs of $\Gamma=0$, $0.1$, and $1$, respectively.
All the three curves apparently diverge to infinity at $\tau=0$.
Moreover,
the curve of $\Gamma=0$ seemingly diverges to infinity at $\tau=2\pi$.}
\label{Figure11}
\end{figure}

\begin{figure}
\begin{center}
\includegraphics{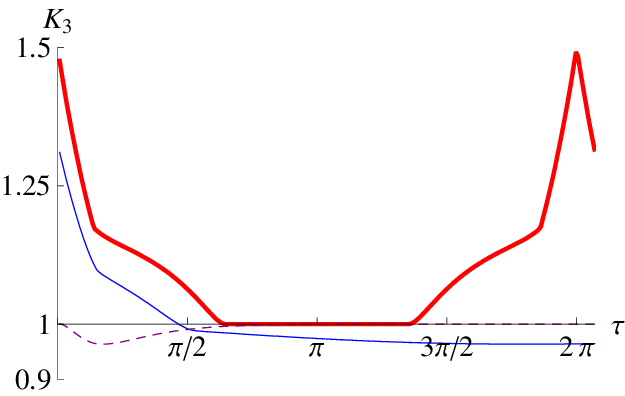}
\end{center}
\caption{Graphs of the maximized $K_{3}$,
with adjusting $\theta$ and $r$, for each $\tau(>0)$
as a function of $\tau$,
where the system is initialized in the cat state.
The thick solid red, thin solid blue, and thin dashed purple curves represent plots of $\Gamma=0$, $0.1$, and $1$, respectively.
As $\Gamma$ increases, amplitudes of the curves diminish.}
\label{Figure12}
\end{figure}

Next, in a similar way to Sect.~\ref{section-initial-coherent-state-numerical-analyses},
we consider the following optimization problem.
We look for the optimum values of $\theta$ and $r$ that maximize $K_{3}$ for given arbitrary $\tau(>0)$.
In Figs.~\ref{Figure09} and \ref{Figure11},
we plot the optimum $\theta$ and $r$ that maximize $K_{3}$ for each $\tau(>0)$ as functions of $\tau$.
In Fig.~\ref{Figure12},
we plot the maximized $K_{3}$ with adjusting $\theta$ and $r$ for each $\tau(>0)$ as a function of $\tau$.
In Figs.~\ref{Figure09}, \ref{Figure11}, and \ref{Figure12},
the thick solid red, thin solid blue, and thin dashed purple curves represent plots of $\Gamma=0$, $0.1$, and $1$, respectively.
In Fig.~\ref{Figure10},
a thick red curve represents the maximized $K_{3}$ for each $\tau(>0)$ and $\Gamma=0$ with adjustments of $\theta$ and $r$ as a function of $\tau$,
and thin blue lines consist of
$\theta=(5\pi/2)-\tau$, $\theta=2\pi-\tau$, $\theta=\pi-\tau$, and $\theta=(\pi/2)-\tau$.

Here,
we focus on Fig.~\ref{Figure10}.
The optimized $\theta$ that makes $K_{3}$ be maximum moves and jumps on the four parallel blue lines.
In the following,
we examine this fact in detail.

From slightly tough calculations,
we obtain
\begin{eqnarray}
\left.\frac{\partial}{\partial\theta}K_{3}(0;\theta,r,\tau)\right|_{\theta=(\pi/2)-\tau}
&=&
\left.\frac{\partial}{\partial\theta}K_{3}(0;\theta,r,\tau)\right|_{\theta=\pi-\tau} \nonumber \\
&=&
0.
\end{eqnarray}
Because $K_{3}(0;\theta,r,\tau)$ is a periodic function about $\theta$ and its period is given by $\pi$,
\\
$\left.(\partial/\partial\theta)K_{3}(0;\theta,r,\tau)\right|_{\theta=2\pi-\tau}
=\left.(\partial/\partial\theta)K_{3}(0;\theta,r,\tau)\right|_{\theta=(5\pi/2)-\tau}=0$ holds obviously.
Thus, in the cases of $\theta=(\pi/2)-\tau$, $\theta=\pi-\tau$, $\theta=2\pi-\tau$, and $\theta=(5\pi/2)-\tau$,
we can expect that $K_{3}(0;\theta,r,\tau)$ takes extreme values.

In Fig.~\ref{Figure10},
$\theta=0$ holds for $2.050\leq\tau\leq 4.250$.
Looking at the curve of $\Gamma=0$ in Fig.~\ref{Figure11},
we can confirm that $r=0$ holds for this range of $\tau$.
Thus,
we can regard $\theta$ as meaningless for $2.050\leq\tau\leq 4.250$.

In a similar fashion to the case where the initial state is given by a coherent state $|\alpha\rangle$,
letting the initial state be the cat state,
we can find strong singularity of $K_{3}$ around $\tau=0$.
Due to the definition of $K_{3}$ in Eq.~(\ref{definition-Leggett-Garg-inequality-0}),
we suppose that $K_{3}(0;\theta,r,0)=1$ seemingly holds.
However,
the maximum value of $K_{3}$ approaches $1.5$ under the limit $\tau\rightarrow +0$ in Fig.~\ref{Figure12}.
From numerical calculations,
we can confirm that the maximum value of $K_{3}$ attains $1.499{\,}02...$ at $\tau=0.001$.
We investigate these facts in detail in the following paragraphs.

Looking at Fig.~\ref{Figure10},
we suppose that the maximized $K_{3}$ is given by $\theta=(\pi/2)-\tau$ around $\tau=0$.
Thus,
we consider a function:
\begin{equation}
g(r,\tau)
=
K_{3}(0;(\pi/2)-\tau,r,\tau).
\end{equation}
We apply the following approximations to $g(r,\tau)$ in the above equation.
Taking the limit $\tau\rightarrow +0$,
we put $\sin\tau\simeq \tau$ and $\cos\tau\simeq 1-\tau^{2}/2$.
Moreover,
we set $r\gg 1$.
Then, we obtain
\begin{eqnarray}
g(r,\tau)
&\simeq&
\frac{\exp(-2r\tau)}{1+\sqrt{e}}
\Biggl\{
[1+\exp(4r\tau)+2\exp(\frac{1}{2}+2r\tau)]
\cos(4r^{2}\tau) \nonumber \\
&&
-
[1+\exp(\frac{1}{2}+16r^{2}\tau^{2})]\cos(8r^{2}\tau)
\Biggr\}.
\label{g_approximation_CAT}
\end{eqnarray}

Being similar to that in Eq.~(\ref{g_approximation}),
the above $g(r,\tau)$ is very intractable,
so that we cannot compute $g(r,\tau)$ for $r\to\infty$ and $\tau\to +0$ with ease.
Thus,
we use the same technique for estimation of the limit as discussed in Sect.~\ref{section-initial-coherent-state-numerical-analyses}.

First, we put $x=r^{2}\tau$ and we assume that $x$ is fixed at a finite value.
Second, putting $x$ at the finite value,
we take the limits, $\tau\to +0$ and $r\to\infty$.
Then, because of $r\tau=x/r\to 0$ and $r^{2}\tau^{2}=x\tau\to 0$,
we obtain the following function for the limit of $g(r,\tau)$:
\begin{equation}
g(x)
=
2\cos 4x-\cos 8x.
\end{equation}
This equation has appeared in Sect.~\ref{section-initial-coherent-state-numerical-analyses} already and $g(x)$ attains the maximum value $3/2$ at $x=\pi/12$.
Thus, we can suppose that the maximum value of $K_{3}(0;(\pi/2)-\tau,r,\tau)$ converges to $3/2$ in the limits, $r\to\infty$ and $\tau\to +0$.

We have verified that $r^{2}\tau\simeq\pi/12$ holds around $\tau\simeq 0$ for $\Gamma=0$ numerically.
To be more precise,
we have obtained that $r^{2}\tau=\mbox{Const.}(\pi/12)$ with $0.9934\leq\mbox{Const.}\leq 0.9991$ for $\tau=0.001\times n$ and $n=1,2,...,8$.
Furthermore, we have made sure that $\mbox{Const.}$ increases and approaches unity as $\tau$ gets closer to zero.
Therefore,
we can guess that the maximum value of $K_{3}$ converges to $3/2$ in the limit $\tau\to +0$.

\section{\label{section-coherent-CAT-states-numerical-analyses}
Comparisons of the LGIs
for cases where the systems are initially prepared in the coherent and cat states}

\begin{figure}
\begin{center}
\includegraphics{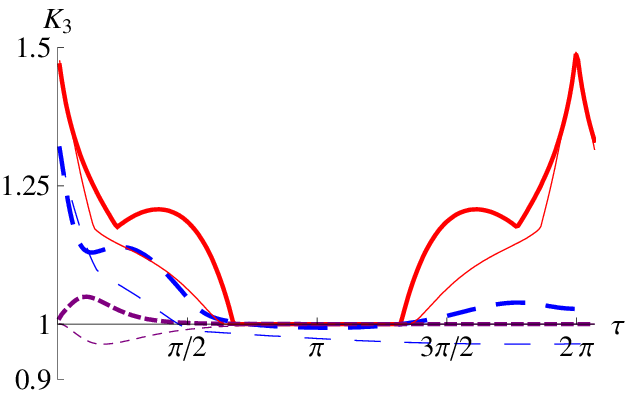}
\end{center}
\caption{Graphs of the maximized $K_{3}$ as a function of $\tau$.
The thick and thin curves represent plots of $K_{3}$ for the systems initially set
in the coherent state $|\alpha\rangle$ and the cat state $(|\alpha\rangle+|-\alpha\rangle)$ respectively,
where $\alpha=1/2$ and $\omega=1$.
The solid red, long dashed blue, and short dashed purple curves represent plots of $\Gamma=0$, $0.1$, $1$, respectively.
We notice that the coherent state shows a larger violation of the LGI than the cat state for $\tau$ in a specific range.}
\label{Figure13}
\end{figure}

\begin{figure}
\begin{center}
\includegraphics{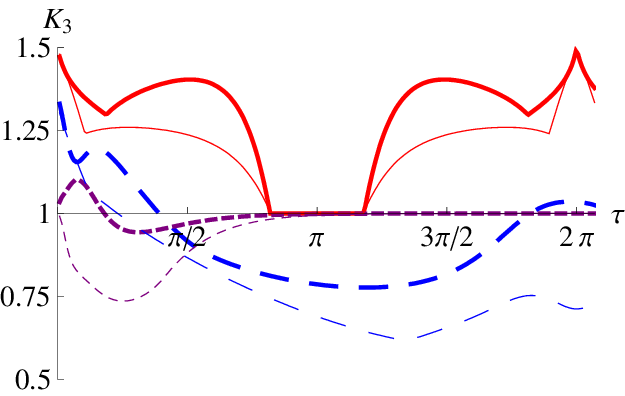}
\end{center}
\caption{Graphs of the maximized $K_{3}$ as a function of $\tau$.
The thick and thin curves represent plots of $K_{3}$ for the systems prepared initially
in the coherent state $|\alpha\rangle$ and the cat state $(|\alpha\rangle+|-\alpha\rangle)$ respectively,
where $\alpha=1$ and $\omega=1$.
The solid red, long dashed blue, and short dashed purple curves represent plots of $\Gamma=0$, $0.1$, and $1$, respectively.
We become aware that the coherent state reveals a larger violation of the LGI than the cat state for $\tau$ in a specific range.}
\label{Figure14}
\end{figure}

In Fig.~\ref{Figure13},
we plot the maximized $K_{3}$ as a function of $\tau$ with $\alpha=1/2$ and $\omega=1$.
The thick and thin curves represent plots for the systems initialized in the coherent and cat states, respectively.
The solid red, long dashed blue, and short dashed purple curves represent graphs of $\Gamma=0$, $0.1$, and $1$, respectively.
In Fig.~\ref{Figure14}, the same graphs are drawn as Fig.~\ref{Figure13} but $\alpha=1$.

We focus on the curves of $\Gamma=0$ in Fig.~\ref{Figure13}.
In the ranges of $0.175\leq \tau\leq 2.125$ and $4.175\leq\tau\leq 6.125$,
the value of the maximized $K_{3}$ for the coherent state is larger than that for the cat state.
By contrast,
in the ranges of $0\leq\tau<0.175$ and $6.125<\tau\leq 2\pi$,
the value of the maximized $K_{3}$ for the cat state is larger than that for the coherent state.
Except for the above ranges,
the values of the maximized $K_{3}$ are equal to each other for both the states.

In Fig.~\ref{Figure14} for $\alpha=1$,
we can observe the same facts.
We concentrate on the curves of $\Gamma=0$.
In the ranges of $0.1\leq \tau\leq 2.55$ and $3.725\leq\tau\leq 6.175$,
the values of the maximized $K_{3}$ for the coherent state is larger than that for the cat state.
Contrastingly,
in the ranges of $0\leq\tau<0.1$ and $6.175<\tau\leq 2\pi$,
the value of the maximized $K_{3}$ for the cat state is larger than that for the coherent state.
Except for the above ranges,
the values of the maximized $K_{3}$ for both the states are equal to each other.

In Figs.~\ref{Figure13} and \ref{Figure14},
for both $\Gamma=0.1$ and $\Gamma=1$,
the value of the maximized $K_{3}$ for coherent state is larger than that for the cat state
in a wide range of $\tau$.

From the above results,
we can conclude that the coherent state exhibits a characteristic of quantum nature more strongly than the cat state
in the specific ranges of $\tau$.

\section{\label{section-discussion}Discussion}
In the current paper,
we demonstrate that the coherent state shows a characteristic of the quantum nature more intensely than the cat state
for the specific ranges of the time difference concerning to the violation of the LGI.
In Ref.~\cite{Chevalier2009},
it has been already pointed out that the coherent state is able to violate the LGI.
In Ref.~\cite{Chevalier2009},
Chevalier {\it et al}. constructed the LGI by using the Mach-Zehnder interferometer for the measurements.
Chevalier {\it et al}. argued how to evaluate the LGI
by injecting the coherent light into the Mach-Zehnder interferometer and applying the negative measurement, that was so-called interaction-free measurement, to it.
In contrast,
Ref.~\cite{Thenabadu2019} showed that the cat state was able to violate the LGI.

In general,
the coherent state has the balanced minimum uncertainty $\Delta X=\Delta P=1/\sqrt{2}$ and it is regarded as one of the most classical-like states.
By contrast, Refs.~\cite{Schleich1991,Buzek1992} mentioned that the cat state was able to exhibit sub-Poisson photon statistics.
From these viewpoints,
we can regard the coherent and cat states as pseudoclassical and nonclassical, respectively.
However, as shown in Sect.~\ref{section-coherent-CAT-states-numerical-analyses},
the violation of the Leggett-Garg inequality of the coherent state is stronger than that of the cat state for time differences of $\tau$ in some ranges.
This fact suggests to us that the violation of the Leggett-Garg inequality can be a witness of non-classicality of wave functions but does not work as a quantitative measure of it.

It is certain that the violation disproves the macroscopic realism of the physical system.
However,
we cannot define the concept of the macroscopicity explicitly.
In actual fact,
Ref.~\cite{Kofler2007} showed that a particular time evolution with coarse-grained measurements caused the macrorealism in a quantum system.
In Ref.~\cite{Moreira2015}, Moreira {\it et al}. considered the macroscopic realism to be a model dependent notion and provided a toy model
in which the invasiveness was controlled by physical parameters.
Because of these circumstances,
we have not obtained quantitative measure of the macrorealism yet.

It is possible that
a choice of observables affects the degree of non-macroscopicity revealed by the violation of the LGI.
In the present paper,
we choose the displaced parity operators for the measurements of the boson system
in the LGI.
We can suppose that this choice lets the coherent state exhibit a characteristic of the quantum nature more strongly than the cat state.

In Ref.~\cite{Ku2020}, it was reported that two- and four-qubit cat states violated the LGI
but a six-qubit cat state did not violate the bound of clumsy-macrorealistic.
(The experimental solution of the clumsy loophole,
in other words clumsy measurement process inducing invasive one and causing a violation rather than quantum effect,
was addressed by Refs.~\cite{Wilde2012,Knee2016}.)
In the current paper,
we cannot determine what kind of quantity of the macroscopic realism the displaced parity operators reflect.
We may be able to let the cat state exhibit a characteristic of quantum nature more strongly than the coherent state
for the violation of the LGI
by opting the other operators for the measurements,
rather than the displaced parity operators.

In order to realize the tests the current paper considers in a laboratory,
we have to perform measurements of quantum states using the displaced parity operators without destroying them.
In other words, quantum nondemolition measurements with the displaced parity operators are essential.
Reference~\cite{Birrittella2021} has reviewed measurements with the parity operators comprehensively.
In Ref.~\cite{Englert1993}, an experimental proposition to detect the parity of the field of one specified mode in a high-$Q$ resonator is explained.
Reference~\cite{Cohen2014} reported experiments for photon-number parity measurements of coherent states
using a photon-number resolving detector and a polarization version of the Mach-Zehnder interferometer.
However, in these experiments, quantum states were destroyed after the observations and the quantum nondemolition measurements were not executed.
The quantum nondemolition measurement with the displaced parity operators is one of the most difficult challenges
in the field of experimental quantum optics.

Recently, as another path to demonstrate a quantum nondemolition measurement on a bosonic mode,
opto-electro-mechanical and nanomechanical systems
have been examined theoretically.
In Ref.~\cite{Lambert2011},
Lambert
{\it et al}. showed that an unambiguous violation of the LGI was given using an opto-electo-mechanical system
with an additional circuit-QED measurement device.
In Ref.~\cite{Johansson2014},
Johansson {\it et al}. considered how to generate entangled states with a multimode nanomechanical resonator and observe
violations of the Bell inequality.
Although these set-ups do not utilize the displaced parity operators,
they give us some suggestions for a realization of the quantum nondemolition measurements of the LGI.

In Figs.~\ref{Figure13} and \ref{Figure14},
we become aware that $K_{3}\neq 1$ holds in the limit $\tau\to +0$ for $\Gamma=0$.
This phenomenon can be observed for a positive decay rate, that is $\Gamma>0$, as well.
It is a novel discovery that the displaced parity operators reveal the strong singularity in the LGI in the limit $\tau\rightarrow +0$.
Referring to Refs.~\cite{Emary2014,Huelga1995},
in general,
$K_{3}=1$ holds at $\tau=0$ for a two-level system
using projection operators as observables.
We can attribute the singularity found in the current paper to the fact that the dimension of the Hilbert space of the system is infinite.

In the current paper,
we examine the LGI for a single boson mode that lies on an infinite dimensional Hilbert space.
In contrast, there are some works concerning the LGI for a multi-level system and an ensemble of qubits.
In Ref.~\cite{Budroni2014}, Budroni and Emary showed that $K_{3}>3/2$ was able to hold for an $N$-level system
such as a large spin with projection operators measuring the spin in the $z$ direction.
It is proved in general that the maximum value of $K_{3}$ is equal to $3/2$ for a two-level system \cite{Leggett1985},
so that their results are interesting.
In Ref.~\cite{Lambert2016}, Lambert {\it et al}. investigated the violation of the LGI for a large ensemble of qubits and showed the following.
When a parity of the projection of the spin in the $z$ direction was chosen as the dichotomic variable for measurement
like Ref.~\cite{Budroni2014},
the violation of the LGI occurred at $\tau$ that approached zero ($\tau\to +0$)
as the number of qubits $N$ became larger.
This observation matches our result of Figs.~\ref{Figure13} and \ref{Figure14}
that the maximized $K_{3}$ appears at $\tau=0$.
However,
the maximized $K_{3}$ shrank to unity as $N$ became larger in Ref.~\cite{Lambert2016}
although $K_{3}$ of Figs.~\ref{Figure13} and \ref{Figure14} attain $3/2$ at $\tau=0$.
We can suppose that our optimization of measurement operators causes this difference.

In the current paper, under the optimization of the displaced parity operators,
we obtain the restriction $K_{3}\leq 3/2$.
In Ref.~\cite{Budroni2013}, it was proven thoroughly that $K_{3}$ must be equal to or less than $3/2$
if the observables $\Pi_{+}$ and $\Pi_{-}$ are projection operators onto eigenspaces of $Q=\pm1$.
Hence, in the case of the current paper,
the relation $K_{3}\leq 3/2$ is valid.
Contrastingly, Ref.~\cite{Budroni2014} showed examples which demonstrated $K_{3}>3/2$.

In the present paper,
we study the boson system coupled to the zero-temperature environment.
The simplest method for analysing time evolution of a boson system that interacts with a thermal reservoir is
solving the master equation with the perturbative approach,
for instance, low temperature expansion.
However,
calculations of this method tend to be complicated,
so that it is not practical.
In Ref.~\cite{Friedenberger2017},
Friedenberger and Lutz derived time evolution of a qubit coupled to the thermal reservoir by using the quantum regression theorem.
Because this approach gives us a clear perspective for solving the time evolution of the qubit,
we may apply it to a problem of the thermal boson system, as well.

\appendix

\section{\label{section-appendix-A}
The mathematical forms used in Sect.~\ref{section-LG-equation-for-coherent-state}}
An explicit form of $w_{1\pm}(\tau)$,
the time evolution of $w_{1\pm}(0)$ given by Eqs.~(\ref{w1plus0-0}) and (\ref{w1minus0-0}), is written down as follows:
\begin{eqnarray}
w_{1\pm}(\tau)
&=&
\frac{1}{4}
\Biggl\{
|\alpha e^{-i\Omega\tau}\rangle\langle \alpha e^{-i\Omega\tau}| \nonumber \\
&&
\pm
\exp\{
\beta^{*}\alpha-\beta\alpha^{*}
-(1/2)[|2\beta-\alpha|^{2}+|\alpha|^{2}-2(2\beta-\alpha)\alpha^{*}](1-e^{-2\Gamma\tau})
\} \nonumber \\
&&
\times
|(2\beta-\alpha)e^{-i\Omega\tau}\rangle\langle\alpha e^{-i\Omega\tau}| \nonumber \\
&&
\pm
(\mbox{the hermitian conjugate of the above term}) \nonumber \\
&&
+
|(2\beta-\alpha)e^{-i\Omega\tau}\rangle\langle (2\beta-\alpha)e^{-i\Omega\tau}|
\Biggr\}.
\label{formula-w1pmtau}
\end{eqnarray}

Explicit forms of $p_{1\pm,2+}$ and $p_{1\pm,2-}$,
the probabilities that $O_{2}=1$ and $O_{2}=-1$ are observed with the measurement on the above $w_{1\pm}(\tau)$ at time $t_{2}$ respectively,
are given by
\begin{eqnarray}
p_{1\pm,2+}
&=&
\mbox{Tr}
[\Pi^{(+)}(\beta)w_{1\pm}(\tau)] \nonumber \\
&=&
\frac{1}{4}
\Biggl\{
\exp(-|\alpha e^{-i\Omega\tau}-\beta|^{2})
\cosh|\alpha e^{-i\Omega\tau}-\beta|^{2} \nonumber \\
&&
\pm
2\mbox{Re}
\Biggr[
\exp\{
\beta^{*}\alpha-\beta\alpha^{*}
-(1/2)[|2\beta-\alpha|^{2}+|\alpha|^{2}-2(2\beta-\alpha)\alpha^{*}](1-e^{-2\Gamma\tau}) \nonumber \\
&&
-\beta(\beta^{*}-\alpha^{*})\exp(i\Omega^{*}\tau)+\beta^{*}(\beta-\alpha)\exp(-i\Omega\tau) \nonumber \\
&&
-(1/2)[|\alpha e^{-i\Omega\tau}-\beta|^{2}+|(2\beta-\alpha)e^{-i\Omega\tau}-\beta|^{2}]
\} \nonumber \\
&&
\times
\cosh[(\alpha^{*}\exp(i\Omega^{*}\tau)-\beta^{*})
((2\beta-\alpha)\exp(-i\Omega\tau)-\beta)]
\Biggl] \nonumber \\
&&
+
\exp[-|(2\beta-\alpha)e^{-i\Omega\tau}-\beta|^{2}]
\cosh|(2\beta-\alpha)e^{-i\Omega\tau}-\beta|^{2}
\Biggr\},
\label{p1pm2+}
\end{eqnarray}
\begin{eqnarray}
p_{1\pm,2-}
&=&
\mbox{Tr}
[\Pi^{(-)}(\beta)w_{1\pm}(\tau)] \nonumber \\
&=&
\frac{1}{4}
\Biggl\{
\exp(-|\alpha e^{-i\Omega\tau}-\beta|^{2})
\sinh|\alpha e^{-i\Omega\tau}-\beta|^{2} \nonumber \\
&&
\pm
2\mbox{Re}
\Biggr[
\exp\{
\beta^{*}\alpha-\beta\alpha^{*}
-(1/2)[|2\beta-\alpha|^{2}+|\alpha|^{2}-2(2\beta-\alpha)\alpha^{*}](1-e^{-2\Gamma\tau}) \nonumber \\
&&
-\beta(\beta^{*}-\alpha^{*})\exp(i\Omega^{*}\tau)+\beta^{*}(\beta-\alpha)\exp(-i\Omega\tau) \nonumber \\
&&
-(1/2)[|\alpha e^{-i\Omega\tau}-\beta|^{2}+|(2\beta-\alpha)e^{-i\Omega\tau}-\beta|^{2}]
\} \nonumber \\
&&
\times
\sinh[(\alpha^{*}\exp(i\Omega^{*}\tau)-\beta^{*})
((2\beta-\alpha)\exp(-i\Omega\tau)-\beta)]
\Biggl] \nonumber \\
&&
+
\exp[-|(2\beta-\alpha)e^{-i\Omega\tau}-\beta|^{2}]
\sinh|(2\beta-\alpha)e^{-i\Omega\tau}-\beta|^{2}
\Biggr\}.
\label{p1pm2-}
\end{eqnarray}

\section{\label{section-appendix-B}
The mathematical forms used in Sect.~\ref{section-LG-equation-for-CAT-state}}
Explicit forms of $\{K^{(j)}(0):j=1,2,4\}$,
$\{L^{(j)}(0):j=1,2,3,4\}$, and
$\{M^{(j)}(0):j=1,2,4\}$
appearing in Eq.~(\ref{PipKLMPip-definition}) are given by
\begin{eqnarray}
K^{(1)}(0)
&=&
|\alpha\rangle\langle\alpha|, \nonumber \\
K^{(2)}(0)
&=&
\exp(\beta\alpha^{*}-\beta^{*}\alpha)|\alpha\rangle\langle -\alpha+2\beta|, \nonumber \\
K^{(4)}(0)
&=&
|-\alpha+2\beta\rangle\langle -\alpha+2\beta|,
\label{K-0-definition}
\end{eqnarray}
\begin{eqnarray}
L^{(1)}(0)
&=&
|\alpha\rangle\langle -\alpha|, \nonumber \\
L^{(2)}(0)
&=&
\exp(-\beta\alpha^{*}+\beta^{*}\alpha)|\alpha\rangle\langle \alpha+2\beta|, \nonumber \\
L^{(3)}(0)
&=&
\exp(-\beta\alpha^{*}+\beta^{*}\alpha)|-\alpha+2\beta\rangle\langle -\alpha|, \nonumber \\
L^{(4)}(0)
&=&
\exp[2(-\beta\alpha^{*}+\beta^{*}\alpha)]|-\alpha+2\beta\rangle\langle \alpha+2\beta|,
\label{L-0-definition}
\end{eqnarray}
\begin{eqnarray}
M^{(1)}(0)
&=&
|-\alpha\rangle\langle -\alpha|, \nonumber \\
M^{(2)}(0)
&=&
\exp(-\beta\alpha^{*}+\beta^{*}\alpha)|-\alpha\rangle\langle \alpha+2\beta|, \nonumber \\
M^{(4)}(0)
&=&
|\alpha+2\beta\rangle\langle \alpha+2\beta|.
\label{M-0-definition}
\end{eqnarray}

Because of Eq.~(\ref{time-evolution-alpha-beta-0}),
time evolution from time $t_{1}=0$ to time $t_{2}=\tau$ of the above operators,
$\{K^{(j)}(\tau):j=1,2,4\}$,
$\{L^{(j)}(\tau):j=1,2,3,4\}$, and
$\{M^{(j)}(\tau):j=1,2,4\}$,
are described in the forms,
\begin{eqnarray}
K^{(1)}(\tau)
&=&
|\alpha \exp(-i\Omega\tau)\rangle\langle\alpha \exp(-i\Omega\tau)|, \nonumber \\
K^{(2)}(\tau)
&=&
\exp\{\beta\alpha^{*}-\beta^{*}\alpha \nonumber \\
&&
-(1/2)[|\alpha|^{2}+|-\alpha+2\beta|^{2}-2\alpha(-\alpha^{*}+2\beta^{*})]
[1-\exp(-2\Gamma\tau)]\} \nonumber \\
&&
\times
|\alpha \exp(-i\Omega\tau)\rangle\langle(-\alpha+2\beta) \exp(-i\Omega\tau)|, \nonumber \\
K^{(4)}(\tau)
&=&
|(-\alpha+2\beta) \exp(-i\Omega\tau)\rangle\langle(-\alpha+2\beta) \exp(-i\Omega\tau)|,
\label{K-tau-definition}
\end{eqnarray}
\begin{eqnarray}
L^{(1)}(\tau)
&=&
\exp\{-2|\alpha|^{2}[1-\exp(-2\Gamma\tau)]\}
|\alpha \exp(-i\Omega\tau)\rangle\langle -\alpha \exp(-i\Omega\tau)|, \nonumber \\
L^{(2)}(\tau)
&=&
\exp\{-\beta\alpha^{*}+\beta^{*}\alpha
-(1/2)[|\alpha|^{2}+|\alpha+2\beta|^{2}-2\alpha(\alpha^{*}+2\beta^{*})]
[1-\exp(-2\Gamma\tau)]\} \nonumber \\
&&
\times
|\alpha \exp(-i\Omega\tau)\rangle\langle (\alpha+2\beta)\exp(-i\Omega\tau)|, \nonumber \\
L^{(3)}(\tau)
&=&
\exp\{-\beta\alpha^{*}+\beta^{*}\alpha \nonumber \\
&&
-(1/2)[|-\alpha+2\beta|^{2}+|\alpha|^{2}+2(-\alpha+2\beta)\alpha^{*}]
[1-\exp(-2\Gamma\tau)]\} \nonumber \\
&&
\times
|(-\alpha+2\beta)\exp(-i\Omega\tau)\rangle\langle -\alpha\exp(-i\Omega\tau)|, \nonumber \\
L^{(4)}(\tau)
&=&
\exp\{2(-\beta\alpha^{*}+\beta^{*}\alpha) \nonumber \\
&&
-(1/2)[|-\alpha+2\beta|^{2}+|\alpha+2\beta|^{2}
-2(-\alpha+2\beta)(\alpha^{*}+2\beta^{*})][1-\exp(-2\Gamma\tau)]\} \nonumber \\
&&
\times
|(-\alpha+2\beta)\exp(-i\Omega\tau)\rangle\langle (\alpha+2\beta)\exp(-i\Omega\tau)|,
\label{L-tau-definition}
\end{eqnarray}
\begin{eqnarray}
M^{(1)}(\tau)
&=&
|-\alpha \exp(-i\Omega\tau)\rangle\langle -\alpha \exp(-i\Omega\tau)|, \nonumber \\
M^{(2)}(\tau)
&=&
\exp\{-\beta\alpha^{*}+\beta^{*}\alpha
-(1/2)[|\alpha|^{2}+|\alpha+2\beta|^{2}+2\alpha(\alpha^{*}+2\beta^{*})]
[1-\exp(-2\Gamma\tau)]\} \nonumber \\
&&
\times
|-\alpha \exp(-i\Omega\tau)\rangle\langle (\alpha+2\beta)\exp(-i\Omega\tau)|, \nonumber \\
M^{(4)}(\tau)
&=&
|(\alpha+2\beta)\exp(-i\Omega\tau)\rangle\langle (\alpha+2\beta)\exp(-i\Omega\tau)|.
\label{M-tau-definition}
\end{eqnarray}

A mathematically rigorous form of $p_{1\pm,2+}$,
the probability that $O_{2}=1$ is obtained with the measurement on $w_{1\pm}(\tau)$ at time $t_{2}$,
is given by
\begin{eqnarray}
p_{1\pm,2+}
&=&
(1/4)q(\alpha)^{-1} \nonumber \\
&&
\times
\Bigl(
\mbox{Tr}[K^{(1)}(\tau)\Pi^{(+)}(\beta)]
\pm
2\mbox{Re}
\{
\mbox{Tr}[K^{(2)}(\tau)\Pi^{(+)}(\beta)]
\}
+\mbox{Tr}[K^{(4)}(\tau)\Pi^{(+)}(\beta)] \nonumber \\
&&
+
2\mbox{Re}
\{
\mbox{Tr}[L^{(1)}(\tau)\Pi^{(+)}(\beta)]
\pm
\mbox{Tr}[L^{(2)}(\tau)\Pi^{(+)}(\beta)]
\pm
\mbox{Tr}[L^{(3)}(\tau)\Pi^{(+)}(\beta)] \nonumber \\
&&
+
\mbox{Tr}[L^{(4)}(\tau)\Pi^{(+)}(\beta)]
\} \nonumber \\
&&
+
\mbox{Tr}[M^{(1)}(\tau)\Pi^{(+)}(\beta)]
\pm
2\mbox{Re}
\{
\mbox{Tr}[M^{(2)}(\tau)\Pi^{(+)}(\beta)]
\}
+\mbox{Tr}[M^{(4)}(\tau)\Pi^{(+)}(\beta)]
\Bigr), \nonumber \\
\label{p1pm2p}
\end{eqnarray}
where explicit forms of
$\{\mbox{Tr}[K^{(j)}(\tau)\Pi^{(+)}(\beta)]:j=1,2,4\}$,
$\{\mbox{Tr}[L^{(j)}(\tau)\Pi^{(+)}(\beta)]:j=1,2,3,4\}$, and
$\{\mbox{Tr}[M^{(j)}(\tau)\Pi^{(+)}(\beta)]:j=1,2,4\}$
are written as
\begin{eqnarray}
\mbox{Tr}[K^{(1)}(\tau)\Pi^{(+)}(\beta)]
&=&
\exp[-|\alpha\exp(-i\Omega\tau)-\beta|^{2}]\cosh|\alpha\exp(-i\Omega\tau)-\beta|^{2}, \nonumber \\
\mbox{Tr}[K^{(2)}(\tau)\Pi^{(+)}(\beta)]
&=&
\exp
\{
\beta\alpha^{*}-\beta^{*}\alpha \nonumber \\
&&
-(1/2)[|\alpha|^{2}+|-\alpha+2\beta|^{2}-2\alpha(-\alpha^{*}+2\beta^{*})]
[1-\exp(-2\Gamma\tau)] \nonumber \\
&&
-[\beta\alpha^{*}\exp(i\Omega^{*}\tau)-\beta^{*}\alpha\exp(-i\Omega\tau)]
+2i|\beta|^{2}\sin\omega\tau\exp(-\Gamma\tau) \nonumber \\
&&
-(1/2)[|\alpha\exp(-i\Omega\tau)-\beta|^{2}
+|(-\alpha+2\beta)\exp(-i\Omega\tau)-\beta|^{2}]
\} \nonumber \\
&&
\times
\cosh\{[(-\alpha^{*}+2\beta^{*})\exp(i\Omega^{*}\tau)-\beta^{*}]
[\alpha\exp(-i\Omega\tau)-\beta]\}, \nonumber \\
\mbox{Tr}[K^{(4)}(\tau)\Pi^{(+)}(\beta)]
&=&
\exp[-|(-\alpha+2\beta)\exp(-i\Omega\tau)-\beta|^{2}] \nonumber \\
&&
\times
\cosh|(-\alpha+2\beta)\exp(-i\Omega\tau)-\beta|^{2},
\label{K-formula}
\end{eqnarray}
\begin{eqnarray}
\mbox{Tr}[L^{(1)}(\tau)\Pi^{(+)}(\beta)]
&=&
\exp
\{
-2|\alpha|^{2}[1-\exp(-2\Gamma\tau)]
-[\beta\alpha^{*}\exp(i\Omega^{*}\tau)-\beta^{*}\alpha\exp(-i\Omega\tau)] \nonumber \\
&&
-(1/2)[|\alpha\exp(-i\Omega\tau)-\beta|^{2}+|\alpha\exp(-i\Omega\tau)+\beta|^{2}]
\} \nonumber \\
&&
\times
\cosh\{-[\alpha^{*}\exp(i\Omega^{*}\tau)+\beta^{*}][\alpha\exp(-i\Omega\tau)-\beta]\}, \nonumber \\
\mbox{Tr}[L^{(2)}(\tau)\Pi^{(+)}(\beta)]
&=&
\exp
\{
-\beta\alpha^{*}+\beta^{*}\alpha \nonumber \\
&&
-(1/2)[|\alpha|^{2}+|\alpha+2\beta|^{2}-2\alpha(\alpha^{*}+2\beta^{*})]
[1-\exp(-2\Gamma\tau)] \nonumber \\
&&
+2i|\beta|^{2}\sin\omega\tau\exp(-\Gamma\tau) \nonumber \\
&&
-(1/2)[|\alpha\exp(-i\Omega\tau)-\beta|^{2}
+|(\alpha+2\beta)\exp(-i\Omega\tau)-\beta|^{2}]
\} \nonumber \\
&&
\times
\cosh\{[(\alpha^{*}+2\beta^{*})\exp(i\Omega^{*}\tau)-\beta^{*}]
[\alpha\exp(-i\Omega\tau)-\beta]\}, \nonumber \\
\mbox{Tr}[L^{(3)}(\tau)\Pi^{(+)}(\beta)]
&=&
\exp
\{
-\beta\alpha^{*}+\beta^{*}\alpha \nonumber \\
&&
-(1/2)[|-\alpha+2\beta|^{2}+|\alpha|^{2}+2(-\alpha+2\beta)\alpha^{*}]
[1-\exp(-2\Gamma\tau)] \nonumber \\
&&
-2i|\beta|^{2}\sin\omega\tau\exp(-\Gamma\tau) \nonumber \\
&&
-(1/2)[|(-\alpha+2\beta)\exp(-i\Omega\tau)-\beta|^{2}
+|\alpha\exp(-i\Omega\tau)+\beta|^{2}]
\} \nonumber \\
&&
\times
\cosh\{-[\alpha^{*}\exp(i\Omega^{*}\tau)+\beta^{*}]
[(-\alpha+2\beta)\exp(-i\Omega\tau)-\beta]\}, \nonumber \\
\mbox{Tr}[L^{(4)}(\tau)\Pi^{(+)}(\beta)]
&=&
\exp
\{
2(-\beta\alpha^{*}+\beta^{*}\alpha) \nonumber \\
&&
-(1/2)[|-\alpha+2\beta|^{2}+|\alpha+2\beta|^{2}
-2(-\alpha+2\beta)(\alpha^{*}+2\beta^{*})] \nonumber \\
&&
\times
[1-\exp(-2\Gamma\tau)] \nonumber \\
&&
-[-\beta\alpha^{*}\exp(i\Omega^{*}\tau)+\beta^{*}\alpha\exp(-i\Omega\tau)] \nonumber \\
&&
-(1/2)[|(-\alpha+2\beta)\exp(-i\Omega\tau)-\beta|^{2} \nonumber \\
&&
+|(\alpha+2\beta)\exp(-i\Omega\tau)-\beta|^{2}]
\} \nonumber \\
&&
\times
\cosh\{[(\alpha^{*}+2\beta^{*})\exp(i\Omega^{*}\tau)-\beta^{*}]
[(-\alpha+2\beta)\exp(-i\Omega\tau)-\beta]\}, \nonumber \\
\label{L-formula}
\end{eqnarray}
\begin{eqnarray}
\mbox{Tr}[M^{(1)}(\tau)\Pi^{(+)}(\beta)]
&=&
\exp[-|\alpha\exp(-i\Omega\tau)+\beta|^{2}]
\cosh|\alpha\exp(-i\Omega\tau)+\beta|^{2}, \nonumber \\
\mbox{Tr}[M^{(2)}(\tau)\Pi^{(+)}(\beta)]
&=&
\exp
\{
-\beta\alpha^{*}+\beta^{*}\alpha
-(1/2)[|\alpha|^{2}+|\alpha+2\beta|^{2}+2\alpha(\alpha^{*}+2\beta^{*})] \nonumber \\
&&
\times
[1-\exp(-2\Gamma\tau)] \nonumber \\
&&
+[\beta\alpha^{*}\exp(i\Omega^{*}\tau)-\beta^{*}\alpha\exp(-i\Omega\tau)]
+2i|\beta|^{2}\sin\omega\tau\exp(-\Gamma\tau) \nonumber \\
&&
-(1/2)[|\alpha\exp(-i\Omega\tau)+\beta|^{2}
+|(\alpha+2\beta)\exp(-i\Omega\tau)-\beta|^{2}]
\} \nonumber \\
&&
\times
\cosh\{-[(\alpha^{*}+2\beta^{*})\exp(i\Omega^{*}\tau)-\beta^{*}]
[\alpha\exp(-i\Omega\tau)+\beta]\}, \nonumber \\
\mbox{Tr}[M^{(4)}(\tau)\Pi^{(+)}(\beta)]
&=&
\exp[-|(\alpha+2\beta)\exp(-i\Omega\tau)-\beta|^{2}] \nonumber \\
&&
\times
\cosh|(\alpha+2\beta)\exp(-i\Omega\tau)-\beta|^{2}.
\label{M-formula}
\end{eqnarray}

An explicit form of $p_{1\pm,2-}$,
the probability that $O_{2}=-1$ is obtained with the measurement on $w_{1\pm}(\tau)$ at time $t_{2}$,
is described in the form,
\begin{eqnarray}
p_{1\pm,2-}
&=&
(1/4)q(\alpha)^{-1} \nonumber \\
&&
\times
\Bigl(
\mbox{Tr}[K^{(1)}(\tau)\Pi^{(-)}(\beta)]
\pm
2\mbox{Re}
\{
\mbox{Tr}[K^{(2)}(\tau)\Pi^{(-)}(\beta)]
\}
+\mbox{Tr}[K^{(4)}(\tau)\Pi^{(-)}(\beta)] \nonumber \\
&&
+
2\mbox{Re}
\{
\mbox{Tr}[L^{(1)}(\tau)\Pi^{(-)}(\beta)]
\pm
\mbox{Tr}[L^{(2)}(\tau)\Pi^{(-)}(\beta)]
\pm
\mbox{Tr}[L^{(3)}(\tau)\Pi^{(-)}(\beta)] \nonumber \\
&&
+
\mbox{Tr}[L^{(4)}(\tau)\Pi^{(-)}(\beta)]
\} \nonumber \\
&&
+
\mbox{Tr}[M^{(1)}(\tau)\Pi^{(-)}(\beta)]
\pm
2\mbox{Re}
\{
\mbox{Tr}[M^{(2)}(\tau)\Pi^{(-)}(\beta)]
\}
+\mbox{Tr}[M^{(4)}(\tau)\Pi^{(-)}(\beta)]
\Bigr), \nonumber \\
\label{p1pm2m}
\end{eqnarray}
where
$\{\mbox{Tr}[K^{(j)}(\tau)\Pi^{(-)}(\beta)]:j=1,2,4\}$,
$\{\mbox{Tr}[L^{(j)}(\tau)\Pi^{(-)}(\beta)]:j=1,2,3,4\}$, and
\\
$\{\mbox{Tr}[M^{(j)}(\tau)\Pi^{(-)}(\beta)]:j=1,2,4\}$
are given by
$\{\mbox{Tr}[K^{(j)}(\tau)\Pi^{(+)}(\beta)]:j=1,2,4\}$,
\\
$\{\mbox{Tr}[L^{(j)}(\tau)\Pi^{(+)}(\beta)]:j=1,2,3,4\}$, and
$\{\mbox{Tr}[M^{(j)}(\tau)\Pi^{(+)}(\beta)]:j=1,2,4\}$
in Eqs.~(\ref{K-formula}), (\ref{L-formula}), and (\ref{M-formula})
with replacing $\cosh$ with $\sinh$,
that is substitution of hyperbolic sines for hyperbolic cosines.

Explicit forms of $p_{2\pm,3+}$ and $p_{2\pm,3-}$,
the probabilities that $O_{3}=1$ and $O_{3}=-1$ are obtained with the measurement on $w_{2\pm}(2\tau)$ at $t_{3}$ respectively,
are given by
\begin{eqnarray}
p_{2\pm,3+}
&=&
(1/4)q(\alpha)^{-1} \nonumber \\
&&
\times
\Bigl(
\mbox{Tr}[\tilde{K}^{(1)}(\tau)\Pi^{(+)}(\beta)]
\pm
2\mbox{Re}
\{
\mbox{Tr}[\tilde{K}^{(2)}(\tau)\Pi^{(+)}(\beta)]
\}
+\mbox{Tr}[\tilde{K}^{(4)}(\tau)\Pi^{(+)}(\beta)] \nonumber \\
&&
+
2\exp[-2|\alpha|^{2}(1-e^{-2\Gamma\tau})] \nonumber \\
&&
\times
\mbox{Re}
\{
\mbox{Tr}[\tilde{L}^{(1)}(\tau)\Pi^{(+)}(\beta)]
\pm
\mbox{Tr}[\tilde{L}^{(2)}(\tau)\Pi^{(+)}(\beta)]
\pm
\mbox{Tr}[\tilde{L}^{(3)}(\tau)\Pi^{(+)}(\beta)] \nonumber \\
&&
+
\mbox{Tr}[\tilde{L}^{(4)}(\tau)\Pi^{(+)}(\beta)]
\} \nonumber \\
&&
+
\mbox{Tr}[\tilde{M}^{(1)}(\tau)\Pi^{(+)}(\beta)]
\pm
2\mbox{Re}
\{
\mbox{Tr}[\tilde{M}^{(2)}(\tau)\Pi^{(+)}(\beta)]
\}
+\mbox{Tr}[\tilde{M}^{(4)}(\tau)\Pi^{(+)}(\beta)]
\Bigr), \nonumber \\
\label{p2pm2p-formula}
\end{eqnarray}
\begin{eqnarray}
p_{2\pm,3-}
&=&
(1/4)q(\alpha)^{-1} \nonumber \\
&&
\times
\Bigl(
\mbox{Tr}[\tilde{K}^{(1)}(\tau)\Pi^{(-)}(\beta)]
\pm
2\mbox{Re}
\{
\mbox{Tr}[\tilde{K}^{(2)}(\tau)\Pi^{(-)}(\beta)]
\}
+\mbox{Tr}[\tilde{K}^{(4)}(\tau)\Pi^{(-)}(\beta)] \nonumber \\
&&
+
2\exp[-2|\alpha|^{2}(1-e^{-2\Gamma\tau})] \nonumber \\
&&
\times
\mbox{Re}
\{
\mbox{Tr}[\tilde{L}^{(1)}(\tau)\Pi^{(-)}(\beta)]
\pm
\mbox{Tr}[\tilde{L}^{(2)}(\tau)\Pi^{(-)}(\beta)]
\pm
\mbox{Tr}[\tilde{L}^{(3)}(\tau)\Pi^{(-)}(\beta)] \nonumber \\
&&
+
\mbox{Tr}[\tilde{L}^{(4)}(\tau)\Pi^{(-)}(\beta)]
\} \nonumber \\
&&
+
\mbox{Tr}[\tilde{M}^{(1)}(\tau)\Pi^{(-)}(\beta)]
\pm
2\mbox{Re}
\{
\mbox{Tr}[\tilde{M}^{(2)}(\tau)\Pi^{(-)}(\beta)]
\}
+\mbox{Tr}[\tilde{M}^{(4)}(\tau)\Pi^{(-)}(\beta)]
\Bigr), \nonumber \\
\label{p2pm2m-formula}
\end{eqnarray}
where
\begin{eqnarray}
\mbox{Tr}[\tilde{K}^{(j)}(\tau)\Pi^{(\pm)}(\beta)]
&=&
\left. \mbox{Tr}[K^{(j)}(\tau)\Pi^{(\pm)}(\beta)] \right|_{\alpha\rightarrow \alpha\exp(-i\Omega\tau)}, \nonumber \\
\mbox{Tr}[\tilde{L}^{(j)}(\tau)\Pi^{(\pm)}(\beta)]
&=&
\left. \mbox{Tr}[L^{(j)}(\tau)\Pi^{(\pm)}(\beta)] \right|_{\alpha\rightarrow \alpha\exp(-i\Omega\tau)}, \nonumber \\
\mbox{Tr}[\tilde{M}^{(j)}(\tau)\Pi^{(\pm)}(\beta)]
&=&
\left. \mbox{Tr}[M^{(j)}(\tau)\Pi^{(\pm)}(\beta)] \right|_{\alpha\rightarrow \alpha\exp(-i\Omega\tau)}
\quad
\mbox{for $j=1,2,3,4$}.
\end{eqnarray}

\end{document}